\documentclass[sigconf]{acmart}
\AtBeginDocument{%
  }

\setcopyright{acmlicensed}
\copyrightyear{2027}
\acmYear{2027}
\acmDOI{XXXXXXX.XXXXXXX}
\acmConference[Preprint, In Submission]{Make sure to enter the correct
  conference title from your rights confirmation email}{Publication time}{Publishing venue}
\acmISBN{978-1-4503-XXXX-X/2018/06}

\usepackage{amsmath}
\usepackage{algorithm}
\usepackage{algorithmic}
\usepackage{graphicx}
\usepackage{booktabs}
\usepackage{multirow}

\usepackage{booktabs}       
\usepackage{graphicx}
\usepackage{xcolor}
\usepackage{placeins}

\usepackage[table]{xcolor}
\usepackage{array}
\usepackage{graphicx}

\newcommand{\gcell}[2]{%
  \ifcase#1\relax
    \cellcolor[HTML]{E8F6EC}\textcolor[HTML]{06351C}{\textbf{#2}}\or
    \cellcolor[HTML]{C6E8D0}\textcolor[HTML]{06351C}{\textbf{#2}}\or
    \cellcolor[HTML]{8FCF9F}\textcolor[HTML]{06351C}{\textbf{#2}}\or
    \cellcolor[HTML]{58B072}\textcolor[HTML]{06351C}{\textbf{#2}}\or
    \cellcolor[HTML]{2E8B4A}\textcolor{white}{\textbf{#2}}\or
    \cellcolor[HTML]{1B6B38}\textcolor{white}{\textbf{#2}}\or
    \cellcolor[HTML]{0E4D28}\textcolor{white}{\textbf{#2}}\or
    \cellcolor[HTML]{06351C}\textcolor{white}{\textbf{#2}}%
  \fi}
\newcommand{\bcell}[2]{%
  \ifcase#1\relax
    \cellcolor[HTML]{EAF4F8}\textcolor[HTML]{082B43}{\textbf{#2}}\or
    \cellcolor[HTML]{C7E3ED}\textcolor[HTML]{082B43}{\textbf{#2}}\or
    \cellcolor[HTML]{91C6D8}\textcolor[HTML]{082B43}{\textbf{#2}}\or
    \cellcolor[HTML]{5A9FBB}\textcolor[HTML]{082B43}{\textbf{#2}}\or
    \cellcolor[HTML]{367C9D}\textcolor{white}{\textbf{#2}}\or
    \cellcolor[HTML]{205B7A}\textcolor{white}{\textbf{#2}}\or
    \cellcolor[HTML]{123F5A}\textcolor{white}{\textbf{#2}}\or
    \cellcolor[HTML]{082B43}\textcolor{white}{\textbf{#2}}%
  \fi}
\newcommand{\mh}[1]{\shortstack{\bfseries #1}}
\newcolumntype{C}{>{\centering\arraybackslash}p{0.078\textwidth}}

\title{Measured Joules, Learned Routes: Learning to Route for Energy-Efficient LLM Serving}
\author{Muhammad Abdur Rab Siddiqui}
\email{msiddiq6@ualberta.ca}
\affiliation{%
  \institution{University of Alberta}
  \city{Edmonton}
    \state{Alberta}
  \country{Canada}
}

\author{Daniela Rojas}
\email{d2rojas@ucsd.edu}
\affiliation{%
  \institution{University of California San Diego}
  \city{La Jolla}
  \state{California}
  \country{USA}
}

\author{Chen Yang}
\email{cyang529@connect.hkust-gz.edu.cn}
\affiliation{%
  \institution{Hong Kong University of Science and Technology (Guangzhou)}
  \city{Guangzhou}
  \country{China}
}

\author{Wenqi Cui}
\email{wenqicui@nyu.edu}
\affiliation{%
  \institution{New York University}
  \city{New York}
  \state{New York}
  \country{USA}
}

\author{Yuanyuan Shi}
\email{yyshi@ucsd.edu}
\affiliation{%
  \institution{University of California San Diego}
  \city{La Jolla}
  \state{California}
  \country{USA}
}

\author{Yize Chen}
\email{yize.chen@ualberta.ca}
\affiliation{%
  \institution{University of Alberta}
  \city{Edmonton}
      \state{Alberta}
  \country{Canada}
}

\renewcommand{\shortauthors}{Siddiqui et al.}

\begin{document}

\begin{abstract}
Large language models (LLMs) and agentic AI systems are creating rapidly growing inference energy demands as model sizes grow and reasoning trajectories extend. While in practice, many queries do not require the capabilities of the largest available model, and routinely directing such queries to a high-capability model can introduce unnecessary, considerable computation and energy consumption. In this paper, we investigate whether adaptive routing across a heterogeneous pool of LLMs can reduce this energy burden without substantially compromising task performance. We design a language-model-based router that reads in each query and selects an answer model from a fixed candidate pool. The candidate models are first profiled through an offline tournament that records their correctness, latency, power, and GPU energy for each query. Using these measurements, the router is trained through supervised fine-tuning followed by group relative policy optimization (GRPO) with the tailored paradigms. Results demonstrate that learned routing can selectively allocate expensive model capacity based on query contexts, and effectively improve the accuracy–energy trade-off of multi-LLM serving. Across seven benchmark tasks, we also observe a sharp accuracy–energy phase transition among routers, providing practical insights in terms of striking the energy efficiency while still keeping LLM performance.

\end{abstract}

\maketitle

\section{Introduction}
Artificial intelligence (AI), particularly large language models (LLMs), already places substantial demands on computing infrastructure and its electricity supply. Inference is becoming the central part of this energy burden. As deployed models with billions or trillions of parameters repeatedly process requests throughout their operational lifetime, so even modest energy costs per response accumulate at service scale~\cite{oviedo2026energy,chung2026ml}. Longer reasoning outputs and agentic workflows can increase this demand further by generating more tokens
and invoking models multiple times for a single user request~\cite{chung2026joules}. These workloads make inference efficiency an important concern for sustainable AI serving with regard to cutting down energy spent answering queries while preserving task performance.

As current LLMs vary by model sizes, specialized areas, and associated computing demands, model routing offers one approach to this challenge. Instead of assigning every query to the same LLM, in practice, a router can select an answer model from a heterogeneous pool according to the query's requirements and characteristics~\cite{song2025irt, li2026llmrouterbench}. As modern AI models differ to a significant extent in their capabilities across reasoning, coding, and knowledge-intensive tasks, it creates opportunities to allocate computational effort selectively~\cite{hu2024routerbench}. Always sticking to a high-capability model can spend unnecessary energy on queries that a smaller model could answer correctly, whereas always using a small model can sacrifice accuracy on more demanding inputs. Query-dependent LLM model selection therefore offers a way to exploit model diversity and adjust the quality-cost trade-off~\cite{ong2025routellm,jitkrittum2026universal, hu2024routerbench}.

However, making routing effective and tailored for energy reduction presents a few challenges. Indeed, the cost of a model invocation depends on both models and queries. Input and output lengths, hardware, and serving configurations affect energy consumption, so current practices like parameter counts, API prices, and fixed model-level costs do not fully expose the joules consumed by the underlying generation task~\cite{wilkins2024hybrid,chung2026ml}. This issue becomes especially important in agentic LLM systems, where one user request may trigger multiple model calls for planning, tool use, verification, and revision~\cite{sres}. Moreover, an energy-centric router shall be able to judge and evaluate whether a candidate will answer correctly before observing its output. Reducing energy by systematically choosing weaker models can undermine the service's purpose, and the useful routing requires learning when additional model capacity is warranted. Another concern is routing itself consumes computation, while most of recent works on router design may not distinguish the selected model's consumption from the overhead of making and executing that decision~\cite{ong2025routellm, jitkrittum2026universal}.

\begin{figure}[]
  \centering
  \includegraphics[width=0.45\textwidth]{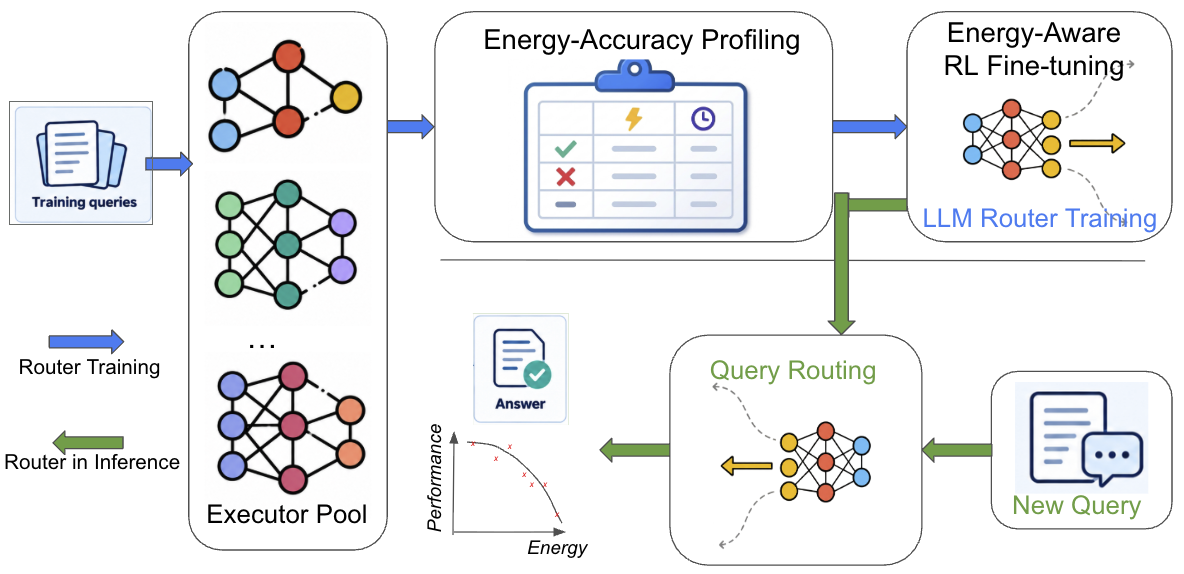}
  \caption{Overview of LLM-based routing learned from measured inference outcomes and RL feedbacks.
  We collect and construct offline profiling dataset from LLM queries to record
  correctness, GPU energy, and latency to train a small LLM routing controller through SFT and GRPO; At serving time, the intended controller selects one candidate from
  the fixed executer pool, which achieves improved tradeoff between LLM serving energy and performance.}
  \label{fig:overview}
\end{figure}

In this work, we study whether a tailored small language model can learn the unique energy-aware routing policy from measured AI serving outcomes. To achieve this router training, we first conduct an offline tournament in which every candidate answer model processes every training question, producing logs of correctness, latency, power, and GPU energy. These logs define an oracle route and a gated reward, where we propose that incorrect answers receive a fixed penalty, while correct answers receive a reward reduced by the selected model's within-query energy rank. Built upon recent advancements in agentic post-training paradigms~\cite{luo2025agent, jiang2025verltool} together with this novel profiling data, the routing controller is trained through supervised fine-tuning (SFT) followed by group relative policy optimization (GRPO)~\cite{shao2024deepseekmath}, including a variant regularized toward the SFT policy. Rather than guided to output complete final answers, based on query contexts, we design the training paradigm to let the router only output the routing decision, which is both lightweight and highly generalizable after finetuning on the small energy-aware datasets. During training and evaluation, model outcomes are replayed from the tournament logs, so answer models do not need to be invoked repeatedly. Figure \ref{fig:overview} illustrates our proposed router training and implementation workflow. 

This work makes three main contributions. First, we formulate LLM model routing around measured GPU energy rather than model size or monetary price alone. Second, we illustrate a practical way of profiling LLM energy and performance during inference, and develop a compact LLM-based routing controller trained through an SFT-to-RL pipeline under a gated accuracy-energy reward. Third, we evaluate proposed and classical model routers on a nine-model pool across a diverse held-out mixture of seven benchmarks. Results indicate a sharp accuracy-energy transition near 60\% accuracy, where LLM models below this region can remain relatively inexpensive, whereas higher accuracy requires substantially more frequent use of capable models. Our proposed policy trained via the mixed KL-GRPO scheme reaches similar accuracy as UniRoute~\cite{jitkrittum2026universal}
, while consistently using about $23\%$ less mean answer energy
($1692\,\mathrm{J}$ vs.\ $2192\,\mathrm{J}$). Ablation studies also indicate interesting observations regarding reward selection, model collapse, and tradeoff between AI performance and energy footprints. We will open-source the proposed router after reviewing period.

\subsection{Related Work}
\label{sec:related}

\subsubsection{Energy-efficient LLM inference}
Due to the surging energy demand associated with AI usage, research start to investigate the power and energy impact of AI training and AI serving~\cite{chung2026ml, li2024unseen}. Inference energy costs are alarming as agentic tasks involve complicated reasoning loops, long reasoning chains, and tendency of using larger sizes of LLMs~\cite{luccioni2024power}. Researchers have identified hardware selection~\cite{wilkins2024hybrid}, AI model training~\cite{you2023zeus}, LLM model caching and serving strategies~\cite{wang2025storellm, he2025freesh}, preemptive power management~\cite{patel2024characterizing} as opportunities of energy reduction.
Prior work has measured substantial variation in LLM inference energy across tasks, models, and deployment configurations~\cite{chung2026ml, oviedo2026energy}.
Such measurement literature shows that energy depends on model size, batching, hardware, precision, and generation length.
For instance, most routers shown in Table~\ref{tab:final_results} still optimize accuracy, embedding similarity, or API-style cost proxies rather than measured GPU energy on the serving stack.
GreenServ~\cite{ziller2026greenserv} is the main exception among our baselines
because it explicitly trades accuracy against normalized inference cost in the
bandit reward.
We evaluate classical and learned routers on the same held-out mix with logged
GPU energy, power, and latency, and we study how reward shaping and RL reshape
the accuracy--energy frontier over a multi-model local pool.

\subsubsection{Routers for AI Serving}
A growing line of work considers practical serving systems, and treats model selection as a first-class inference problem instead of fixing one backend for every query. Most of these routers are designed to optimize efficiency~\cite{jitkrittum2026universal}, while outputs with least token outputs may not be directly translated to energy consumption.
RouterBench formalizes this setting with a large
offline benchmark of per-model outcomes and popular routing baselines~\cite{hu2024routerbench}.
Among them, $k$-nearest-neighbour routing scores each candidate from the error
rates of models on embedding neighbours of the test prompt. UniRoute~\cite{jitkrittum2026universal} represents each LLM through its errors on
representative prompt clusters, so a model can be scored and routed to even when it
was not in the original training pool. Smoothie~\cite{guha2024smoothie} takes a label-free route.
It embeds each candidate's prompt--output pair and estimates per-sample quality
without correctness labels on the routing split. RouteLLM~\cite{ong2025routellm} learns a binary strong--weak router from preference data and thresholds a win probability to trade cost against quality. In our design, we adapt that rule to a strong--weak pair from the local
candidate pool rather than the paper's API checkpoints. GreenServ~\cite{ziller2026greenserv} targets energy-aware serving. It models routing as a contextual bandit and uses LinUCB~\cite{li2010contextual} is  a classical bandit-based approach that maintains a disjoint linear reward model per arm and explores with an uncertainty bonus.
GreenServ's reward mixes accuracy with normalized inference cost, which is close
in spirit to our gated accuracy--energy objective, while our controller is a
learned language model rather than a hand-built feature vector.

The broader AI serving system's goal is not a single router knob, posing challenges to router design. Indeed, agentic serving has two coupled controls. One is which model answers a query. The other is how much reasoning that call is allowed to spend. Most routers named above decide only the first. ARES~\cite{sres} decides the second, and it does so adaptively.
It chooses how much reasoning effort an agent should spend on the current
query, so harder items can think longer and easier items can stop sooner.
It does not assign one fixed thinking budget to every call.
That split is part of what motivates this work.
A fixed model wastes energy on easy items and still fails hard ones 
and a fixed thinking budget has the same problem inside one model.
ARES is the closest prior statement of that second failure, and it treats
reasoning effort as something the system should choose rather than freeze.
Our controller attacks the first control with a small language model and a
gated accuracy--energy reward on measured GPU logs.
Together the two controls point to the same end goal. Agentic inference should
spend compute only when the task justifies it.

\subsubsection{Agentic RL}

Training the router as a policy rather than only as a classifier connects our
setup to agentic RL for language models. Recent advances have shown effectiveness of RL in post training which go beyond text imitation usually happening in SFT approaches.
AgentLightning~\cite{luo2025agent} is an agent training framework that
wraps rollout generation and policy optimization for multi-step workflows.
Hugging Face TRL~\cite{vonwerra2020trl} is a separate library that implements
GRPO and related on-policy algorithms for language models.
We explored both stacks on the same offline tournament reward.
Our reported SFT, GRPO, and KL-GRPO numbers use TRL, so training, checkpoints,
and held-out evaluation stay in one consistent pipeline.
Agent Lightning experiments served as a cross-check on the same reward design
rather than as a second training path in the main tables. The distinction between supervised fine-tuning (SFT) and on-policy RL matters here~\cite{yu2026dapo}. Prior analyses show that RL can improve exploration and reward shaping beyond imitation of static routing labels, while SFT remains a stable warm start~\cite{chu2025sft}. To that end, in this work we study whether RL on replayed tournament rewards can learn when extra answer-model cost is worth higher accuracy under measured energy constraints.

\section{Problem Formulation}
\label{sec:formulation}
\subsection{Heterogeneous model pool}
For modern AI model serving, it is hard to apply one LLM model fit all inputs. This is due to the fact that it becomes challenging to strike a balance between being accurate on hard items and staying efficient on easy queries. For instance, a small, open-weight model generates few tokens and spends little energy, but it can fail queries which need more model and reasoning capacity.
Whereas a large, reasoning-based model does the reverse, because they may overthink, and can waste significant amount of joules on items a smaller model would have easily answered~\cite{zhao2026let}.

To this end, we study a serving system that selects one answer LLM for each incoming query. Let $\mathcal{M}=\{m_1,\ldots,m_K\}$ denote a fixed pool of $K$ candidate LLMs, and let $\mathcal{A}=\{1,\ldots,K\}$ index the available routes. A query context $x$ contains the question and any metadata provided to the router. The routing policy $\pi_\theta(a\mid x)$ selects a candidate
index $a$, after which model $m_a$ generates an answer $y$. The router makes the model-selection decision; the selected candidate performs the task. Each evaluated query involves one routing decision and one answer-model outcome. Multi-turn recovery and tool use could also be applied to our framework with proper extensions.

The pool of candidate LLMs is heterogeneous by design, and the model card is illustrated in Table \ref{tab:candidate_models}. Candidates differ in model family, parameter count, and reasoning style, from sub-billion Qwen2.5 drafts through mid-size models such as Gemma2-2B, Qwen2.5-7B, and Llama~3.1~8B, up to DeepSeek-R1 8B and 32B. That spread is what makes routing meaningful. These differences create opportunities for query-dependent selection, since a model that is effective for one task need not offer the best accuracy--energy trade-off for another. For example, Figure~\ref{fig:pool_example} shows a GSM8K query answered correctly by
Qwen2.5-0.5B using $82\,\mathrm{J}$ and by DeepSeek-R1-8B using $2864\,\mathrm{J}$, while Llama~3.1-8B answers incorrectly using $638\,\mathrm{J}$.  For this question, though Llama model has larger size, its overthinking and extra steps around equation modeling causes both inefficient and inaccurate inference. Such an example motivates measuring outcomes for each query-energy-model triplet rather than inferring suitability from model size alone. We assume all answer models are open-weight checkpoints served locally through Ollama on that device, so each candidate can be invoked under the same power logger rather than billed as a separate API price.

\begin{figure}[ht]
\centering
\small
\setlength{\fboxsep}{6pt}
\noindent\textbf{GSM8K, question 31.}\;
Gunter is trying to count the jelly beans in a jar. He asks his friends how many they think are in the jar. One says~80. Another says 20 more than half the first one. A third says 25\% more than the first one. What is their average guess?
\vspace{0.6em}

\noindent\fcolorbox{green!50!black}{green!8}{%
\begin{minipage}{0.96\linewidth}
\textbf{Qwen2.5-0.5B}\;\textsc{(correct)}\hfill $82$\,J\;\textbar\;$3.8$\,s\\
\textit{Answer:} 80\\
\textit{Response:} ``First, let's calculate each friend's guess: (1) The first friend guesses 80 jelly beans. (2) The second friend guesses 20 more than half of 80, which is $(80/2)+20=60$. (3) The third friend guesses 25\% more than the first one, which is $1.25\times 80=100$. The average is $(80+60+100)/3=80$.''
\end{minipage}}

\vspace{0.45em}
\noindent\fcolorbox{red!50!black}{red!8}{%
\begin{minipage}{0.96\linewidth}
\textbf{Llama~3.1-8B}\;\textsc{(wrong)}\hfill $638$\,J\;\textbar\;$7.8$\,s\\
\textit{Answer:} 106.67\\
\textit{Response:} ``Let's denote Gunter's friend who said 80 as F1, the second friend as F2, and the third friend as F3. F2 $=(1/2)\cdot$F1$+20$. F3 $=1.25\cdot$F1. Now we can find the average of their guesses by adding up all three guesses and dividing by 3: Average = (F1 + ((1/2) * F1 + 20) + (1.25 * F1))/3 Combine like terms. Average = (F1 + (0.5 * F1) + 20 + (1.25 * F1))/3 Simplify the numerator. Average = (3.75 * F1 + 20)/3
Now we can plug in the value of F1, which is 80. Average = (3.75 * 80 + 20)/3
Multiply 3.75 and 80. Average = (300 + 20)/3 Add 300 and 20. Average = 320/3 Divide 320 by 3. Average = 106.67''
\end{minipage}}

\vspace{0.45em}
\noindent\fcolorbox{green!50!black}{green!8}{%
\begin{minipage}{0.96\linewidth}
\textbf{DeepSeek-R1-8B}\;\textsc{(correct)}\hfill $2{,}864$\,J\;\textbar\;$22.8$\,s\\
\textit{Answer:} 80\\
\textit{Response:} ``The first friend guesses 80 jelly beans. The second friend guesses 20 more than half of the first guess: half of 80 is 40, so $40+20=60$. The third friend guesses 25\% more than the first guess: 25\% of 80 is 20, so $80+20=100$. The average is $(80+60+100)/3=80$.''
\end{minipage}}
\caption{Same benchmark query from GSM8K, different profilings of three local models. Responses are truncated tournament logs; energy and duration are measured effective GPU cost for that answer call.}
\label{fig:pool_example}
\end{figure}

\subsection{Recorded outcomes and energy measurement}
\label{sec:outcome_measurement}

For query $i$ and candidate $a$, let $y_{ia}$ be the recorded answer and
$y_i^\star$ its reference. We denote correctness by
\begin{equation}
 C_{ia}=g_{b(i)}(y_{ia},y_i^\star)\in\{0,1\},
 \label{eq:correctness}
\end{equation}
where $g_{b(i)}$ is the evaluator for benchmark $b(i)$. This notation covers
task-specific answer matching and code-test evaluation without requiring
exact and literal string equality. Let $E_{ia}$, $L_{ia}$, and $\bar P_{ia}$ denote
the each answer call $i$'s incremental GPU energy in joules, wall-clock duration in
seconds, and mean incremental GPU power in watts, respectively.

The candidate models are served locally through Ollama on NVIDIA
A100-SXM4-80GB GPUs with tensor parallelism. During profiling, the logger polls board power through
NVML using \texttt{pynvml} at a $200\,\mathrm{ms}$ interval and subtracts
an idle-power baseline. We adopt \texttt{nvidia-ml-py} bindings, which read the same device counters as \texttt{nvidia-smi}. Let $P_{ia}^{\mathrm{board}}(t)$ denote the board power
trace and $P_{ia}^{\mathrm{idle}}$ the baseline associated with the query call $i$,
the measurement quantities of power, energy, and duration are modeled as
\begin{align}
 P_{ia}^{\mathrm{inc}}(t)
   &=P_{ia}^{\mathrm{board}}(t)-P_{ia}^{\mathrm{idle}};
   \label{eq:incremental_power}\\
 E_{ia}
   &=\int_{t_{ia}^{\mathrm{start}}}^{t_{ia}^{\mathrm{end}}}
       P_{ia}^{\mathrm{inc}}(t)\,\mathrm{d}t; 
   \label{eq:call_energy}\\
 L_{ia}&=t_{ia}^{\mathrm{end}}-t_{ia}^{\mathrm{start}},
 \qquad \bar P_{ia}=\frac{E_{ia}}{L_{ia}}.
 \label{eq:call_duration_power}
\end{align}
The energy integral is estimated from the sampled trace. Note these quantities describe idle-subtracted GPU consumption within the measured query call interval, which is the core energy  associated with AI inference. They do not represent total server or data-center-level energy consumption. 

We first profile AI model's energy and performance through an offline tournament of mild size, which records the outcomes of the candidate models on the collected questions. With given query input $x_i$, the per-query record is profiled as
\begin{equation}
 \mathcal{D}_i=
 \left(x_i,\{(C_{ia},E_{ia},L_{ia},\bar P_{ia})\}_{a=1}^{K}\right).
 \label{eq:tournament_record}
\end{equation}

These are realized outcomes of the underlying LLM inference generations. Replaying a record reproduces that recorded outcome, rather than estimating the answer model's stochastic behavior or measurement variation. The complete outcome vector provides feedback for all candidates on a training query, although the learned policy selects only one candidate at inference time.

\subsection{Energy-Aware AI serving objectives}
\label{sec:routing_objective}

In this work, we explore the possibilities of meeting LLM service objectives at low energy cost. On a fixed logged dataset $\mathcal{D}$ of $N$ queries, the expected accuracy
and answer energy of a feasible routing policy are
\begin{subequations}
\begin{align}
 \widehat A_{\mathcal{D}}(\theta)
 &=\frac{1}{N}\sum_{i=1}^{N}\sum_{a=1}^{K}
       \pi_\theta(a\mid x_i)\, C_{ia},
       \label{eq:quality_obj}\\
 \widehat E_{\mathcal{D}}(\theta)
 &=\frac{1}{N}\sum_{i=1}^{N}\sum_{a=1}^{K}
       \pi_\theta(a\mid x_i)\, E_{ia}.
       \label{eq:expected_answer_energy}
\end{align}
\end{subequations}

For deterministic routing, these expressions reduce to averages of the selected candidates' outcomes. For sampled decisions, the corresponding sample averages estimate these expectations. Handling of invalid controller outputs is discussed in Section~\ref{sec:agentic-rl}.

One design to handle the desired accuracy-energy trade-off is
\begin{equation}
 \max_\theta\ \widehat A_{\mathcal{D}}(\theta)
 \quad\text{subject to}\quad
 \widehat E_{\mathcal{D}}(\theta)\leq B;
 \label{eq:energy_constraint}
\end{equation}
where $B$ is an average answer-energy budget in joules per query. This
constraint expresses a service goal, not a guarantee made by the training
algorithm. The reported method instead optimizes the gated rank-based
surrogate defined in Section~\ref{sec:reward} and evaluates its resulting
accuracy and measured answer energy. Problem \eqref{eq:energy_constraint} can be treated as a single-step contextual decision problem. A contextual-bandit formulation accommodates either a classical or a text-conditioned neural policy. In addition, our logged training data provide detailed outcomes for every candidate. We therefore distinguish the routing problem, the controller's representation, and the feedback used to train it.

We note with the addition of the router policy $\pi_\theta$, the complete serving energy also depends on the controller and deployment costs, including model loading and residency under the chosen serving configuration. Those costs is measured use the consistent metric to establish end-to-end savings, and they are not supplied by the lookup of $E_{ia}$ alone. More importantly, as LLM's consumed energy are strongly dependent on input queries and output contexts, it requires dedicated design and investigation of the router.

\begin{table}[ht]
\centering
\caption{Candidate language models used in the evaluation, with ARC Challenge~\cite{clark2018think} accuracy and mean effective answer energy per query on all 472 collected questions.}
\label{tab:candidate_models}
\begin{tabular}{lccc}
\toprule
\textbf{Model} & \textbf{Nominal Size} & \textbf{ARC Acc.} & \textbf{Energy (J)} \\
\midrule
Qwen 2.5      & 0.5B & 0.314 & 66.1 \\
Qwen 2.5      & 1.5B & 0.665 & 84.3 \\
Qwen 2.5      & 3B   & 0.727 & 131.7 \\
Phi-3 Mini    & 3.8B & 0.803 & 170.5 \\
Gemma 2       & 2B   & 0.722 & 56.2 \\
Qwen 2.5      & 7B   & 0.871 & 140.8 \\
Llama 3.1     & 8B   & 0.803 & 196.1 \\
DeepSeek-R1   & 8B   & 0.932 & 2142.2 \\
DeepSeek-R1   & 32B  & 0.939 & 3307.9 \\
\bottomrule
\end{tabular}
\end{table}

\begin{table*}[]
\centering
\caption{Held-out avg.\ results by routing policies.}
\label{tab:final_results}
\begin{tabular}{lccccc}
\toprule
\textbf{Policy} & \textbf{Accuracy} & \textbf{Reward} & \textbf{Duration\ (ms)} & \textbf{Model Energy (J)} & \textbf{Power (W)} \\
\midrule
always\_smallest & 0.189 & - & 3780 & 91.8 & 23.2 \\
always\_medium  & 0.791 & - & 28672 & 4127.9 & 125.9 \\
always\_large   & 0.735 & - & 25706 & 5112.4 & 174.5 \\
K-NN            & 0.790 & - & 23366 & 3697.6 & 121.3 \\
UniRoute        & 0.618 & - & 15371 & 2192.2 & 84.8 \\
Smoothie        & 0.599 & - & 9269 & 1070.6 & 59.4 \\
RouteLLM        & 0.721 & - & 25053 & 4958.6 & 170.3 \\
GreenServ ($\lambda{=}0.5$) & 0.567 & - & 5584 & 277.0 & 41.4 \\
\midrule
SFT ($\lambda_e{=}0.5$) & 0.502 & 0.400 & 7821 & 822.8 & 45.9 \\
KL-GRPO (sft $\lambda_e{=}0.5$, rl $\lambda_e{=}0.4$, step 500)  & 0.591 & 0.686 & 10932 & 1339.8 & 47.1 \\
KL-GRPO (sft $\lambda_e{=}0.5$, rl $\lambda_e{=}0.4$, step 1500) & 0.609 & 0.720 & 13118 & 1691.9 & 55.2 \\
GRPO (sft $\lambda_e{=}0.5$, rl $\lambda_e{=}0.4$, step 750)     & 0.584 & 0.666 & 7942  & 811.8 & 48.5 \\
GRPO (both $\lambda_e{=}0.3$, step 1000)                        & 0.573 & 0.612 & 6002  & 297.0 & 46.8 \\
\bottomrule
\end{tabular}
\end{table*}

\section{Agentic Learning to Route}

\subsection{Router's environment}
\label{sec:agentic-rl}
To fully release the routing controller's effectiveness, we propose to utilize a lightweight, finetuned LLM in a text-conditioned approach. Its role is to interpret a query and select a candidate from the fixed pool, using a smaller LLM than most of the candidate answer models. The
controller prompt contains the question like in Figure~\ref{fig:task_model_mix}, a fixed legend mapping labels to candidate indices, and an optional difficulty field when available. Recorded correctness and energy are regarded as training feedback rather than inputs supplied to the controller when selecting a model.

Let $p_\theta(o\mid x)$ denote the probability of a controller completion $o$ given query context $x$, and let $f(o)$ parse its route label into an index in $\mathcal{A}$. The induced route probabilities are
\begin{equation}
 \pi_\theta(a\mid x)
 =\sum_{o:\,f(o)=a}p_\theta(o\mid x).
 \label{eq:route_marginal}
\end{equation}
This separates probabilities over generated texts from probabilities over candidate models. The reported variants use either a short rationale with a \texttt{Choice: <letter>} field or a choice-only completion. Both formats produce the same model-selection action. An un-parseable output is an invalid action and receives a negative training reward. If multi-turn routing is enabled, the router also sees any prior user-assistant turns on the same item, the current turn number, the model chosen last time, and whether that attempt was correct. The router can output Stop when a satisfactory answer already exists and does not require to route to candidate LLMs. In all reported experiments we use a single routing turn per question, so each item produces one letter choice. Probability on invalid outputs is separate from the valid-route probabilities in Eq.~\eqref{eq:route_marginal}.

The tournament contains approximately $3,000$ collected questions from MMLU, GSM8K, BBH, ARC Challenge, HellaSwag, MATH-500, and HumanEval. The reported training and held-out subsets contain $2050$ and $913$ questions, respectively; benchmark counts are given in Table~\ref{tab:train-test-split} in Appendix. The training subset supplies oracle labels and policy rewards. We use Qwen2.5-1.5B-Instruct as a text-conditioned routing controller, and find it achieve desired post-training behaviors with respect to routing decisions. During policy optimization, the controller generates new completions, but answer-model outcomes are retrieved from the fixed training records. Thus, router sampling is on-policy with respect to the rollout controller, while its environmental feedback is replayed.

\subsection{Router reward design}
\label{sec:reward}

To ensure in the RL post-training, the router can effectively evaluate the input queries' contexts and make decisions, we come up with a ranking-based reward for energy, which is in the same scale as of performance reward. Such a reward combines task correctness with preferences over inference costs (e.g., energy, task completion time and etc). For query $i$ and candidate $a$, $C_{ia}$ denotes correctness, $L_{ia}$ denotes answer latency in seconds, and $\bar P_{ia}$
denotes mean incremental GPU power in watts. The measured answer energy $E_{ia}$ is expressed in joules.

We derive a bounded energy penalty from the ascending rank $k_{ia}$ of $E_{ia}$ among the $K$ candidates for the same query:
\begin{equation}
 z_{ia}=\frac{k_{ia}-1}{K-1}\in[0,1].
 \label{eq:energy_rank}
\end{equation}
With distinct energies, zero corresponds to the cheapest candidate and one to the most expensive. For example, energies of $85$, $95$, $280$, and $3800\,\mathrm{J}$ produce penalties of $0$, $1/3$, $2/3$, and $1$.  $z_{ia}$ denotes rank throughout. It is distinct from energy $E_{ia}$. Ranking bounds the training penalty but discards the magnitude of differences in joules.

Let $\boldsymbol{\lambda}=(\lambda_t,\lambda_e,\lambda_p)$ denote
nonnegative cost coefficients associated with session duration, energy rank, and average GPU power. The general gated reward is
\begin{equation}
 r^{\mathrm{gen}}_{ia}(\boldsymbol{\lambda})=
 \begin{cases}
 R_{\mathrm{correct}}-\lambda_t L_{ia}
       -\lambda_e z_{ia}-\lambda_p\bar P_{ia}, & C_{ia}=1,\\
 -R_{\mathrm{wrong}}, & C_{ia}=0.
 \end{cases}
 \label{eq:general_gated_reward}
\end{equation}

Latency, energy, and power describe related but different aspects of a session call. The same energy can be consumed over different durations and at different mean power levels. Thus, separate latency and power coefficients can express preferences over those quantities, but an average-power penalty alone does not enforce an instantaneous device or rack power limit. We distinguish this general reward form from the configuration used in the reported experiments as follows. 

\paragraph{Reported experimental configuration.}
In our paper, all reported reward configurations set $\lambda_t=\lambda_p=0$ to focus on energy analysis, with
$R_{\mathrm{correct}}=2$ and $R_{\mathrm{wrong}}=1$. The active training
reward therefore reduces to
\begin{equation}
 r_{ia}(\lambda_e)
 =r^{\mathrm{gen}}_{ia}(0,\lambda_e,0)
 =\begin{cases}
 2-\lambda_e z_{ia}, & C_{ia}=1,\\
 -1, & C_{ia}=0.
 \end{cases}
 \label{eq:gated_reward}
\end{equation}
Latency and mean GPU power are consequently measured and reported but do not contribute to the optimization signal in these runs. The oracle and policy-optimization descriptions below use this specialization. Invalid controller outputs and unavailable outcomes receive a separate negative
penalty rather than the valid-route score above. Equivalently, the active reward can be expressed as
\begin{equation}
 r_{ia}(\lambda_e)=3C_{ia}-1-\lambda_e C_{ia}z_{ia}.
 \label{eq:reward_expansion}
\end{equation}
Our reward design distinguishes the costs of correct routes, while it assigns equal reward to cheap and expensive failures. It is an empirical learning surrogate instead of a guarantee of meeting the joule budget in Eq.~\eqref{eq:energy_constraint}. Evaluation counts measured answer energy for successful and unsuccessful answers alike. Joint optimization with nonzero latency or power weights is not demonstrated by the reported experiments.

\subsection{Oracle construction and supervised initialization}
\label{sec:sft_router}

For each training query, the oracle target is selected from
\begin{equation}
 a_i^\star\in\operatorname*{arg\,max}_{a\in\mathcal{A}} \; r_{ia}(\lambda_e^{\mathrm{SFT}}),
  \label{eq:oracle_route}
\end{equation}
where $\lambda_e^{\mathrm{SFT}}$ is the coefficient used during target
construction. Under the reported positive coefficients
$0<\lambda_e^{\mathrm{SFT}}\leq1$, every correct route receives a larger
reward than every incorrect route. The oracle therefore selects a lowest-energy correct candidate whenever one exists. Changing the positive coefficient within this range does not change that hard target if the logs and tie-breaking rule are fixed. For $\lambda_e^{\mathrm{SFT}}=0$, correct candidates tie, and the draft's accuracy-only configuration selects the largest correct model. When all candidates are wrong, every valid route ties under the reward.

Let $o_i^\star$ be the target completion encoding the selected route in the
chosen output format. Supervised fine-tuning initializes the controller by
minimizing the completion negative log likelihood
    \begin{equation}
 \mathcal{L}_{\mathrm{SFT}}(\theta)
 =-\frac{1}{N_{\mathrm{tr}}}\sum_{i\in\mathcal{I}_{\mathrm{tr}}}
        \log p_\theta(o_i^\star\mid x_i),
 \label{eq:sft_loss}
\end{equation}
where $\mathcal{I}_{\mathrm{tr}}$ indexes the $N_{\mathrm{tr}}$ training
queries. We can include two variations of the target during SFT. In the choice-only variant, the target contains the route field; in the rationale variant, it also contains an explanation. The resulting checkpoint is evaluated as an SFT baseline and initializes policy optimization. SFT learns to imitate target completions, whereas the next stage uses the rewards of the controller's sampled decisions.

\subsection{Group-relative policy optimization}
\label{sec:router_optimization}

We fine-tune the controller using GRPO~\cite{shao2024deepseekmath}, which is based on groupwise comparisons instead of using a separate critic model. For each
training query $i$, a rollout policy $p_{\theta_{\mathrm{old}}}$ samples $G$ completions $\{o_{ig}\}_{g=1}^{G}$. Each completion is parsed and scored
using its selected candidate's logged outcome and coefficient
$\lambda_e^{\mathrm{RL}}$, or the invalid-action penalty. Denote these completion rewards by $r_{ig}$. Their group statistics and advantages are
\begin{align}
 \mu_i&=\frac{1}{G}\sum_{g=1}^{G}r_{ig},
 \qquad
 \sigma_i^2=\frac{1}{G}\sum_{g=1}^{G}(r_{ig}-\mu_i)^2,
 \label{eq:group_statistics}\\
 A_{ig}&=\frac{r_{ig}-\mu_i}{\sigma_i+\delta},
 \label{eq:group_advantage}
\end{align}
where $\delta>0$ stabilizes the denominator. A completion receives positive
advantage when its reward exceeds the group mean. If all group rewards
are equal, all advantages are zero and that group supplies no
reward-driven update.

At the completion level, the clipped policy surrogate is expressed using importance sampling ratio
\begin{equation}
 \rho_{ig}(\theta)=
 \frac{p_\theta(o_{ig}\mid x_i)}
      {p_{\theta_{\mathrm{old}}}(o_{ig}\mid x_i)},
 \label{eq:policy_ratio}
\end{equation}
with the resulting loss function
\begin{small}
    \begin{equation}
 \begin{split}
 \mathcal{L}_{\mathrm{clip}}(\theta)
 =-\mathbb{E}_{i,\,o_{i1:G}\sim p_{\theta_{\mathrm{old}}}}
 \bigg[\frac{1}{G}\sum_{g=1}^{G}
 \min\big\{&\rho_{ig}(\theta)A_{ig},\\[-2pt]
 &\operatorname{clip}(\rho_{ig}(\theta),
              1-\varepsilon,1+\varepsilon)A_{ig}\big\}\bigg],
 \end{split}
 \label{eq:grpo_loss}
\end{equation}
\end{small}
where $\varepsilon$ controls ratio clipping and is distinct from the
advantage stabilizer $\delta$. This surrogate favors higher-reward
completions while limiting the incentive for large policy-ratio changes.

The KL-regularized variant additionally anchors the controller to the SFT
reference distribution $p_{\mathrm{ref}}$. At the distribution level its
regularized objective is
\begin{equation}
 \begin{split}
 \mathcal{L}_{\mathrm{KL\text{-}GRPO}}(\theta)
 =\mathcal{L}_{\mathrm{clip}}(\theta)
 +\beta\,\mathbb{E}_{i}\big[
 D_{\mathrm{KL}}(&p_\theta(\cdot\mid x_i)
                \Vert p_{\mathrm{ref}}(\cdot\mid x_i))\big],
 \end{split}
 \label{eq:kl_grpo_loss}
\end{equation}
with regularization coefficient $\beta\geq0$. This expression concerns controller completions rather than the marginal distribution over model indices in Eq.~\eqref{eq:route_marginal}. Note KL regularization discourages departure
from the reference policy. It does not by itself guarantee diverse routing or prevent concentration on a single candidate. Some configurations center rewards by subtracting a query-specific mean
candidate reward. If the same scalar is subtracted from every reward before
the group normalization in Eq.~\eqref{eq:group_advantage}, it cancels from
the advantages. Such centering is therefore distinct from any additional
reward scaling or clipping used by the trainer.

\subsection{Training configurations and evaluation}
\label{sec:training}

In this paper, we unify and use TRL to conduct main SFT, GRPO, and KL-GRPO post-training~\cite{vonwerra2020trl}. Policy optimization starts from the corresponding SFT checkpoint, and samples multiple controller completions per training prompt. The main reported configuration uses $\lambda_e^{\mathrm{SFT}}=0.5$ and $\lambda_e^{\mathrm{RL}}=0.4$ as we find these hyperparameters give the best overall SFT and RL performances. The experiments also examine matched
coefficients and alternative checkpoints. These coefficients identify the oracle-construction and policy-reward settings, respectively. Their effects on hard targets, sampled rewards, and learned policies should be distinguished. The output format is also part of each configuration, where choice-only and rationale-generating controllers have different generation costs, and should be compared with their matching SFT initializations.

On held-out queries, the router can be treated as the lightweight add-on to inference engines, where it selects a LLM model and the evaluator retrieves that candidate's stored outcome. We report accuracy, mean answer-model energy, latency, power, and model-selection frequencies.
Reward comparisons require a common evaluation coefficient: reward values computed with different $\lambda_e$ are different metrics even when the policies share an SFT initialization. Accuracy and measured joules provide the primary comparison across configurations and baselines.

The replay protocol isolates model-selection behavior on a fixed outcome table. It does not measure live changes in answer generation, model-loading cost, queueing, or concurrent serving. Controller measurements and answer-model measurements must therefore be distinguished when discussing complete serving costs. The evaluation addresses single-turn routing on the profiled pool. Adaptation to new deployment conditions and multi-turn agent workflows would require additional experiments.

\begin{figure*}[]
\centering
\includegraphics[width=\textwidth]{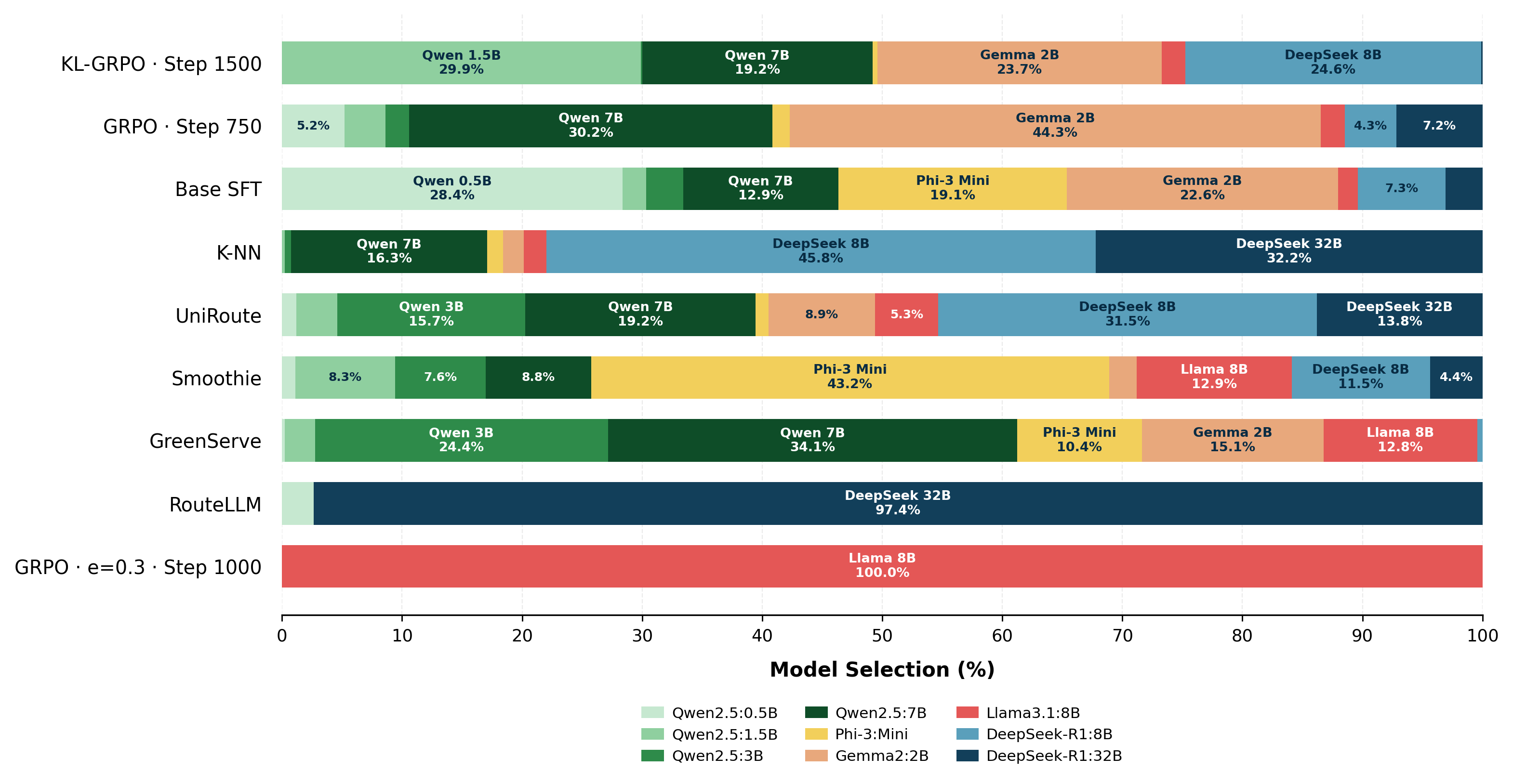}
\caption{Model selection mix on the held-out set.}
\label{fig:results_1}
\end{figure*}

\section{Numerical Studies}

\subsection{Simulation setup}

We evaluate routing policies on a mixed held-out set drawn from seven
standard LLM benchmarks that span knowledge, reasoning, math, and code:
MMLU~\cite{hendrycks2020measuring},
GSM8K~\cite{cobbe2021training},
BBH~\cite{suzgun2023challenging},
ARC Challenge~\cite{clark2018think},
HellaSwag~\cite{zellers2019hellaswag},
MATH-500~\cite{lightman2024let},
and HumanEval~\cite{chen2021evaluating}.
These tasks stress different answer-model costs, from short multiple-choice
items to long generative math and code solutions, which is what makes
energy-aware routing nontrivial on the shared nine-model pool.

Power measurements are based on NVML\footnote{\url{https://docs.nvidia.com/deploy/nvml-api/}}. 
To keep energy and performance comparable across runs, we place all router training and inference on instances of 1x NVIDIA A100-SXM4-$80$,GB GPU. Answer models are served locally with Ollama on that device. The routing controller is trained and run with Hugging Face Transformers and TRL on the same GPU. Offline tournament collection used a separate machine with the same A100-SXM4-$80$,GB specification. During each answer-model generation we poll board power at a $200,\mathrm{ms}$ interval via \texttt{pynvml}, subtract a short idle baseline, and record instantaneous effective power $P(t)$ in watts. Effective energy for the call is $E_{\mathrm{call}}$ as defined in Eq.~\eqref{eq:call_energy}, and latency is the wall-clock duration of the same generation. Table~\ref{tab:final_results} reports mean accuracy together with measured duration, energy, and power summaries.

All methods in Table~\ref{tab:final_results} are scored on the same 913
held-out test questions covering a spectrum of difficulty and categories. Scoring is an offline lookup from pre-logged tournament
outcomes, so no live answer models are called at evaluation time.

K-NN estimates per-model error from embedding neighbours on the train split.
UniRoute matches the prompt to a train K-means cluster and uses per-cluster
error plus a cost term. Smoothie is label-free and scores candidates from
embeddings of $[\mathrm{prompt},\mathrm{output}]$. Unlike the others, Smoothie
uses all logged generations on the test item to decide, so we treat it as an
output-informed diagnostic rather than a deployable pre-generation router.
We still charge only the chosen model's energy. GreenServ is a local LinUCB
adaptation using task, semantic-cluster, and complexity features. Its reward is
$(1-\lambda)\mathrm{Acc}-\lambda\cdot\mathrm{energy}$ and is warmed on train
and frozen before test. It is not the paper's online serving stack.
RouteLLM is a binary router over two fixed answer
models, not a classifier over the full pool.
We set the strong arm to DeepSeek-R1-32B and the weak arm to the smallest
pool model, train a prompt classifier for $P(\text{strong wins}\mid q)$ on
train outcomes, and route to the strong arm when $P\ge\alpha$.
Fixed-model baselines always route every question to one predetermined
answer model (always-smallest, always-medium, or always-large).

Table~\ref{tab:final_results}'s methods and results can be split into three categories. High-accuracy part is K-NN and always-medium,
near $79\%$ at about $3.7$--$4.1\,\mathrm{kJ}$. Always-large and RouteLLM spend even
more and do worse. The mid-accuracy band is UniRoute, Smoothie, and our $\lambda_e{=}0.5{\to}0.4$
GRPO runs, about $58$ to $62\%$ at $0.8$ to $2.2\,\mathrm{kJ}$. The cheap accuracy band is
GreenServ $\lambda{=}0.5$ and matched GRPO $\lambda_e{=}0.3$, about $57\%$ at about
$280$--$300\,\mathrm{J}$. SFT sits below that mid accuracy band. Always-smallest
is cheap in energy but achieves poor accuracy.

\paragraph{\textbf{Reward parameters.}}
We instantiate Eq.~\eqref{eq:gated_reward} with
$R_{\mathrm{correct}}=2$, $R_{\mathrm{wrong}}=1$,
$\lambda_t=\lambda_p=0$, and energy-rank mode for $z_{ia}$ in
Eq.~\eqref{eq:energy_rank}, so the training signal is the gated rank
surrogate rather than raw joules.
Baselines do not train on this gated reward, so the Reward column is blank for
them.
For SFT and TRL the Reward column is therefore
$r_{ia}(\lambda_e)=2-\lambda_e z_{ia}$ when $C_{ia}=1$ and
$r_{ia}(\lambda_e)=-1$ when $C_{ia}=0$.
SFT uses $\lambda_e^{\mathrm{SFT}}=0.5$.
The $0.5{\to}0.4$ runs use $\lambda_e^{\mathrm{RL}}=0.4$.
Matched GRPO uses $\lambda_e=0.3$ at both stages.
Therefore, the Reward column should only be compared among SFT and RL inside that block.
Acc., duration, energy, and power are the columns to compare across the whole
table.

\subsection{Results and discussion} 
\setcounter{dbltopnumber}{3}
\renewcommand{\dbltopfraction}{0.95}
\renewcommand{\dblfloatpagefraction}{0.8}

\begin{figure}[!t]
    \centering
    \includegraphics[width=1\linewidth]{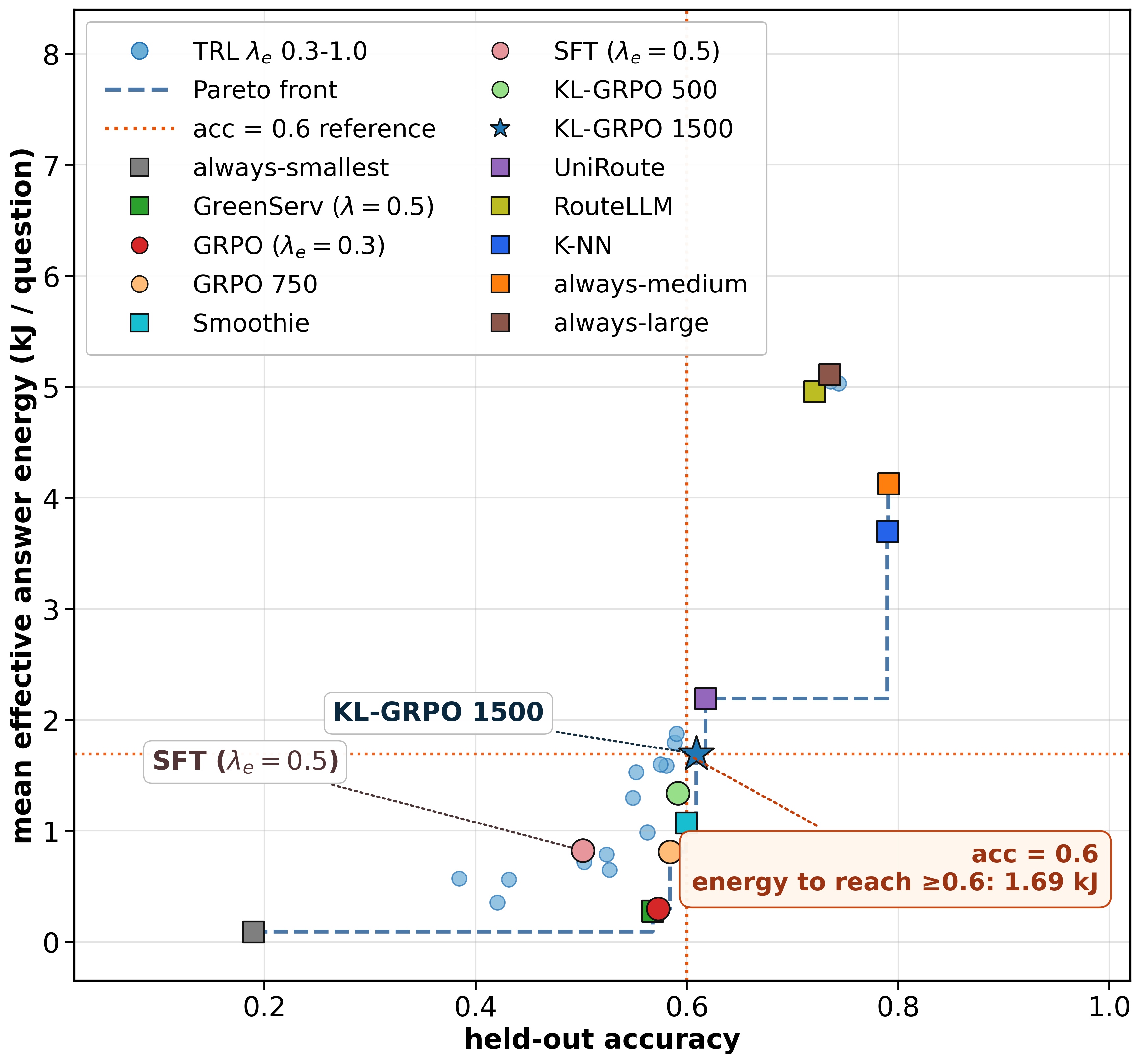}
    \caption{Energy--Accuracy Pareto comparison.}
    \label{fig:pareto_accuracy_energy}
\end{figure}

\paragraph{\textbf{Accuracy--energy trade-off among routers}}
Figure~\ref{fig:pareto_accuracy_energy} plots mean answer energy against
held-out accuracy for the multi-model routers and learned controllers in
Table~\ref{tab:final_results}. Blue markers are TRL sweeps over
$\lambda_e$~\cite{vonwerra2020trl}. Among these methods, policies near
$60\%$ accuracy generally report higher mean answer energy than policies
that remain below that range by relying more on smaller candidate models.
The additional energy is associated with more frequent selection of
higher-capacity models when the query requires it. This comparison is
restricted to the routers and learned controllers under study, and does not
describe every fixed model in the candidate pool. Under that scope, KL-GRPO
step~1500 nearly matches UniRoute ($60.9\%$ vs.\ $61.8\%$) at about $23\%$
lower mean answer energy ($1.69$ vs.\ $2.19\,\mathrm{kJ}$), and GRPO
step~750 is close to Smoothie ($58.4\%$ vs.\ $59.9\%$) at about $24\%$ lower
mean answer energy ($0.81$ vs.\ $1.07\,\mathrm{kJ}$).
    
\paragraph{\textbf{Role of a mid-size fixed arm}}
A single mid-size model can look strong in aggregate. For instance, Always-Qwen2.5-7B
reaches $64.0\%$ at $281.6\,\mathrm{J}$ on the same held-out pool
(Appendix Table~\ref{tab:model_benchmark_profile}), yet it falls to $26.2\%$ on
BBH, so locking every query to 7B can be a weak suite-wide policy even when the
overall mean is high. Our question is not whether one fixed arm can beat a
router on the mean, but whether a controller can allocate capacity across the
pool. Figure~\ref{fig:results_1} shows that the learned and classical routers
still call 7B when it helps. About one third of GreenServ traffic, about
$30\%$ for GRPO step~750, and about $19\%$ for both KL-GRPO step~1500 and
UniRoute use that arm. In that sense 7B is a useful pool member, not a
substitute for routing.
    
\begin{table*}[!tp]
\centering
\setlength{\tabcolsep}{2.4pt}
\renewcommand{\arraystretch}{1.18}
\resizebox{\textwidth}{!}{%
\begin{tabular}{lCCCCCCCCCCC}
\toprule
\mh{ } &
\mh{always\\smallest} &
\mh{GreenServ\\($\lambda{=}0.5$)} &
\mh{Smoothie} &
\mh{SFT\\($\lambda_e{=}0.5$)} &
\mh{GRPO\\(step 750)} &
\mh{KL-GRPO\\(step 1500)} &
\mh{GRPO\\($\lambda_e{=}0.3$)} &
\mh{UniRoute} &
\mh{K-NN} &
\mh{RouteLLM} &
\mh{always\\largest} \\
\midrule
MMLU & \gcell{1}{14.8} & \gcell{4}{51.7} & \gcell{3}{49.0} & \gcell{3}{40.9} & \gcell{3}{38.9} & \gcell{3}{45.6} & \gcell{4}{50.3} & \gcell{4}{58.4} & \gcell{5}{63.1} & \gcell{5}{65.1} & \gcell{5}{65.1} \\
GSM8K & \gcell{2}{27.1} & \gcell{6}{82.6} & \gcell{6}{86.8} & \gcell{3}{38.2} & \gcell{6}{75.7} & \gcell{4}{61.8} & \gcell{6}{81.2} & \gcell{5}{70.8} & \gcell{7}{89.6} & \gcell{7}{89.6} & \gcell{7}{89.6} \\
BBH & \gcell{0}{2.1} & \gcell{1}{15.2} & \gcell{3}{42.8} & \gcell{2}{33.8} & \gcell{3}{40.7} & \gcell{4}{52.4} & \gcell{3}{37.9} & \gcell{1}{19.3} & \gcell{5}{69.7} & \gcell{2}{27.6} & \gcell{2}{27.6} \\
ARC Challenge & \gcell{2}{35.4} & \gcell{6}{77.6} & \gcell{6}{76.2} & \gcell{5}{74.8} & \gcell{6}{76.2} & \gcell{6}{78.2} & \gcell{6}{78.9} & \gcell{6}{87.1} & \gcell{7}{88.4} & \gcell{7}{91.2} & \gcell{7}{92.5} \\
HellaSwag & \gcell{1}{20.4} & \gcell{4}{57.1} & \gcell{4}{53.7} & \gcell{3}{40.8} & \gcell{4}{57.8} & \gcell{4}{58.5} & \gcell{4}{57.8} & \gcell{5}{62.6} & \gcell{5}{72.8} & \gcell{5}{66.7} & \gcell{5}{72.8} \\
Math500 & \gcell{1}{17.2} & \gcell{4}{57.4} & \gcell{4}{62.3} & \gcell{5}{73.0} & \gcell{5}{67.8} & \gcell{5}{73.8} & \gcell{2}{36.9} & \gcell{6}{78.7} & \gcell{6}{86.1} & \gcell{6}{85.2} & \gcell{6}{86.9} \\
HumanEval & \gcell{0}{10.2} & \gcell{4}{54.2} & \gcell{2}{33.9} & \gcell{4}{57.6} & \gcell{3}{47.5} & \gcell{4}{54.2} & \gcell{4}{50.8} & \gcell{4}{52.5} & \gcell{7}{93.2} & \gcell{7}{94.9} & \gcell{7}{94.9} \\
\midrule
\textbf{Overall} & \gcell{1}{18.9} & \gcell{4}{56.7} & \gcell{4}{59.9} & \gcell{4}{50.2} & \gcell{4}{58.4} & \gcell{4}{60.9} & \gcell{4}{57.3} & \gcell{4}{61.8} & \gcell{6}{79.0} & \gcell{5}{72.1} & \gcell{5}{73.5} \\
\bottomrule
\end{tabular}}
\caption{Held-out accuracy (\%) by benchmark and method.\vspace{-10pt}}
\label{tab:accuracy_split}
\end{table*}

\paragraph{\textbf{MATH-500 accounts for most energy}}
Table~\ref{tab:energy_split} breaks energy down by benchmark and shows how
uneven the cost is. Although MATH-500 is only about $13\%$ of the held-out
questions, it accounts for roughly $52$--$57\%$ of the total energy for SFT and
KL-GRPO. For KL-GRPO step~1500 the mean MATH-500 cost is about
$6.6\,\mathrm{kJ}$, while MMLU, GSM8K, ARC, and HellaSwag each sit near
$0.15$--$0.49\,\mathrm{kJ}$. This concentration is why the mid-accuracy rows
still appear expensive in the aggregate table. Collapsed cheap policies reduce
the MATH-500 spend, but Table~\ref{tab:accuracy_split} shows that their
MATH-500 accuracy also drops sharply, from $73.8\%$ under KL-GRPO step~1500
to $36.9\%$ under GRPO~$\lambda_e{=}0.3$.

\paragraph{\textbf{Accuracy is as uneven as energy}}
Table~\ref{tab:accuracy_split} is the matching accuracy view of the same
methods. The Overall row recovers Table~\ref{tab:final_results}, but the
benchmark mix is not uniform. K-NN is the only method that stays high on both
BBH ($69.7\%$) and HumanEval ($93.2\%$); RouteLLM and always-large look strong
in the aggregate table while BBH falls to $27.6\%$. SFT's $50.2\%$ overall hides
a GSM8K collapse to $38.2\%$, whereas GRPO step~750 recovers GSM8K to $75.7\%$
without moving into the high-energy band of Table~\ref{tab:energy_split}.
GreenServ~$\lambda{=}0.5$ and GRPO~$\lambda_e{=}0.3$ remain close overall
($56.7\%$ against $57.3\%$), yet GreenServ keeps MATH-500 at $57.4\%$ while the
collapsed GRPO run does not. KL-GRPO step~1500 is the mixed mid-band policy that
is even on ARC ($78.2\%$) and MATH-500 ($73.8\%$), weaker on MMLU ($45.6\%$),
and still far cheaper than K-NN.

\begin{figure*}[t]
    \centering
    \includegraphics[width=0.92\linewidth]{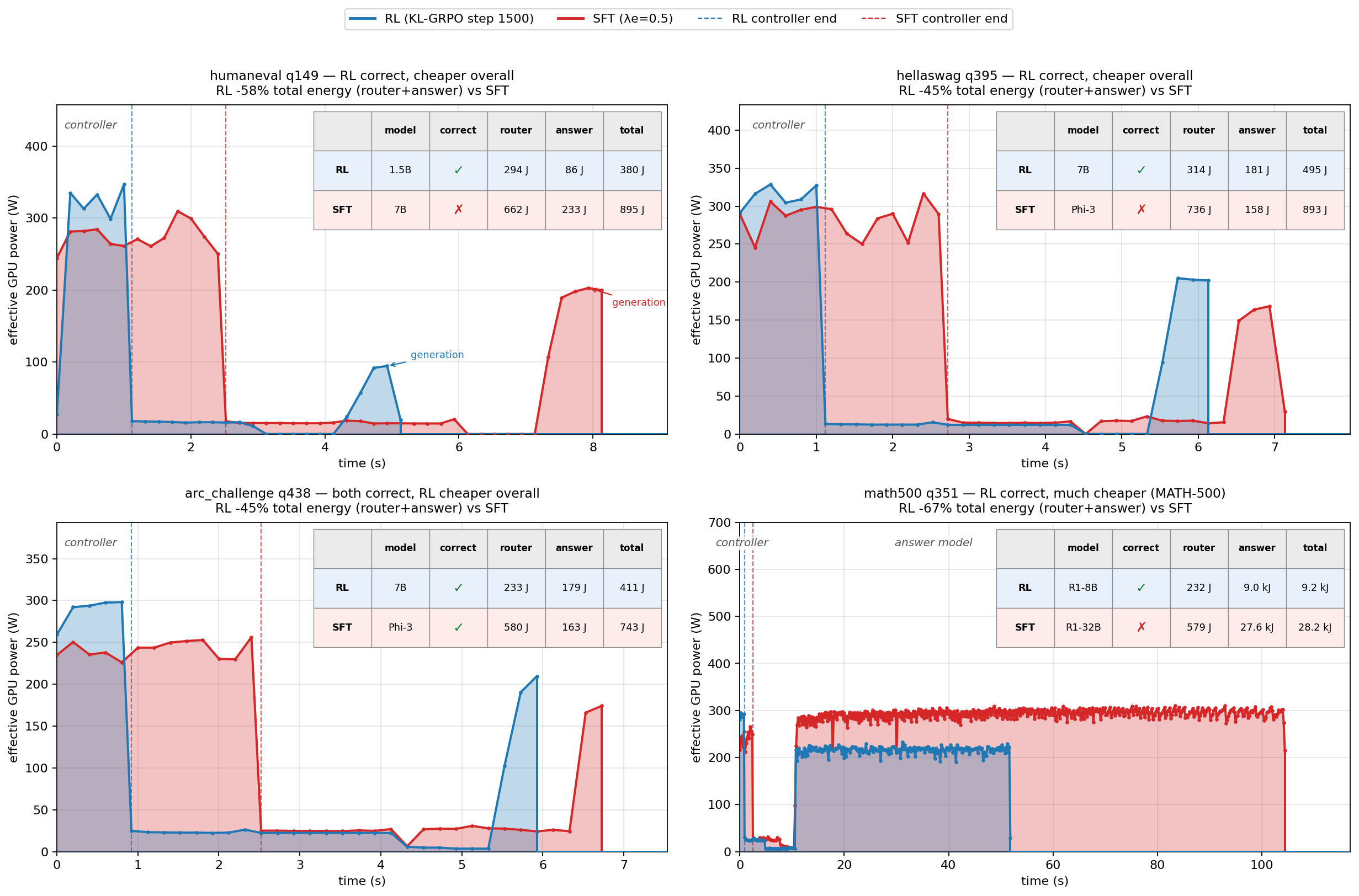}
    \caption{SFT versus KL-GRPO effective GPU power on four held-out questions. Each panel shows the controller decode followed by the answer-model draft, with energy for the router, the answer, and their sum. Title deltas are relative to that total.\vspace{-10pt}}
    \label{fig:placeholder}
\end{figure*}

\paragraph{\textbf{Choice of energy coefficient}}
Our main-table schedule uses $\lambda_e{=}0.5$ at SFT and $\lambda_e{=}0.4$ at
RL. The SFT stage is meant to warm-start an energy-aware hybrid policy that
still under-routes toward cheap drafts. Softening the RL weight to $0.4$ and
adding a KL term to that SFT policy is what lets the controller abandon the
$0.5$B model and move into the mid-accuracy band without jumping to
always-large energy. Setting both stages to $\lambda_e{=}0.3$ instead collapses
onto Llama~3.1~8B on every scored item, with zero selection entropy and the
weak MATH-500 accuracy in Table~\ref{tab:accuracy_split}. Under our hybrid
energy-rank setup, any positive $\lambda_e$ yields the same hard oracle labels
(cheapest correct), so the SFT tags $0.3/0.5/0.7$ share supervision targets.
What changes across these runs is the learned policy and, during RL training, the reward scale. In Table~\ref{tab:final_results}, KL-GRPO step~1500 is the
run among our own policies that reaches above $60\%$ held-out accuracy at the
lowest mean answer energy in that band. The fuller sweep is in
Table~\ref{tab:trl_pareto_checkpoints}.

\paragraph{\textbf{SFT versus RL}}
Reinforcement learning improves the controller from $50.2\%$ accuracy under
SFT to $60.9\%$ at KL-GRPO step~1500. This gain raises mean answer energy from
$0.82$ to $1.69\,\mathrm{kJ}$ because the policy is more willing to pay for a
capable answer model rather than fail cheaply. GRPO step~750 provides a
different operating point, improving accuracy to $58.4\%$ while reducing mean
answer energy to $0.81\,\mathrm{kJ}$. The four examples in
Figure~\ref{fig:placeholder} show that this trade-off is question dependent.
In HumanEval~q149, RL is correct and cheaper overall; in HellaSwag~q395 and
ARC-Challenge~q438 it is also cheaper once router energy is included with the
answer; and in MATH-500~q351, RL avoids an expensive failed
R1-32B call by selecting R1-8B. Thus, the effect of RL is not a uniform
increase in model size or energy, but a change in when additional answer-model
cost is justified for higher accuracy.

\begin{table*}[!tp]
\centering
\setlength{\tabcolsep}{2.4pt}
\renewcommand{\arraystretch}{1.18}
\resizebox{\textwidth}{!}{%
\begin{tabular}{lCCCCCCCCCCC}
\toprule
\mh{ } &
\mh{always\\smallest} &
\mh{GreenServ\\($\lambda{=}0.5$)} &
\mh{Smoothie} &
\mh{SFT\\($\lambda_e{=}0.5$)} &
\mh{GRPO\\(step 750)} &
\mh{KL-GRPO\\(step 1500)} &
\mh{GRPO\\($\lambda_e{=}0.3$)} &
\mh{UniRoute} &
\mh{K-NN} &
\mh{RouteLLM} &
\mh{always\\largest} \\
\midrule
MMLU & \bcell{1}{116.5} & \bcell{2}{213.6} & \bcell{2}{228.8} & \bcell{1}{146.3} & \bcell{1}{123.9} & \bcell{1}{154.2} & \bcell{2}{263.4} & \bcell{6}{2.8k} & \bcell{4}{1.1k} & \bcell{6}{3.9k} & \bcell{6}{3.9k} \\
GSM8K & \bcell{0}{91.1} & \bcell{2}{306.0} & \bcell{5}{1.5k} & \bcell{1}{159.5} & \bcell{3}{389.2} & \bcell{3}{489.1} & \bcell{2}{331.9} & \bcell{3}{572.6} & \bcell{5}{2.3k} & \bcell{5}{2.4k} & \bcell{5}{2.4k} \\
BBH & \bcell{0}{62.7} & \bcell{1}{122.2} & \bcell{5}{1.4k} & \bcell{5}{1.6k} & \bcell{5}{2.3k} & \bcell{6}{2.5k} & \bcell{2}{305.8} & \bcell{2}{234.0} & \bcell{6}{3.6k} & \bcell{6}{4.3k} & \bcell{6}{4.4k} \\
ARC Challenge & \bcell{0}{72.0} & \bcell{1}{157.3} & \bcell{1}{161.8} & \bcell{0}{77.6} & \bcell{1}{129.7} & \bcell{1}{180.1} & \bcell{2}{204.4} & \bcell{5}{1.9k} & \bcell{5}{2.0k} & \bcell{6}{3.2k} & \bcell{6}{3.3k} \\
HellaSwag & \bcell{0}{45.5} & \bcell{1}{127.5} & \bcell{1}{154.3} & \bcell{1}{100.5} & \bcell{3}{382.0} & \bcell{1}{162.3} & \bcell{1}{176.6} & \bcell{5}{1.5k} & \bcell{6}{2.9k} & \bcell{6}{3.4k} & \bcell{6}{3.8k} \\
Math500 & \bcell{1}{167.6} & \bcell{4}{854.0} & \bcell{6}{3.8k} & \bcell{6}{3.5k} & \bcell{5}{2.0k} & \bcell{7}{6.6k} & \bcell{3}{571.4} & \bcell{7}{6.3k} & \bcell{7}{9.6k} & \bcell{7}{8.6k} & \bcell{7}{9.1k} \\
HumanEval & \bcell{1}{110.3} & \bcell{2}{223.7} & \bcell{3}{388.0} & \bcell{2}{312.1} & \bcell{1}{178.3} & \bcell{6}{3.9k} & \bcell{2}{238.0} & \bcell{6}{3.3k} & \bcell{7}{7.8k} & \bcell{7}{16.5k} & \bcell{7}{16.5k} \\
\midrule
\textbf{Overall} & \bcell{0}{91.8} & \bcell{2}{277.0} & \bcell{4}{1.1k} & \bcell{4}{822.8} & \bcell{4}{811.8} & \bcell{5}{1.7k} & \bcell{2}{297.0} & \bcell{5}{2.2k} & \bcell{6}{3.7k} & \bcell{6}{5.0k} & \bcell{6}{5.1k} \\
\bottomrule
\end{tabular}}
\caption{Mean effective answer energy (J) by benchmark and method.}
\label{tab:energy_split}
\end{table*}

The more direct training effect appears in the controller itself, which is the only component updated by SFT and RL. On the held-out evaluation, the majority of SFT controller decisions take about $2.4\,\mathrm{s}$, whereas after KL-GRPO the majority take about $0.8\,\mathrm{s}$, a reduction of more than $50\%$, which is also visible as the shorter controller band in every panel of
Figure~\ref{fig:placeholder}. Because we train only this decoder, that drop is
a change in how the router writes. Indeed, SFT still spends time on a longer reasoning
plus choice generation, while RL concentrates the policy on a short, decisive letter, so the routing step itself becomes cheaper even when the chosen answer model is not. The effective controller power in the plotted traces is consistent across more than thirty checkpoints from various policies collected
on that A100-SXM4-$80$\,GB machine; a later session on the same GPU class produced a different effective draw for otherwise similar decisions, so controller watts follow machine condition more than the training method. Latency, not power, is therefore the reliable signal of how RL changed the
controller in our runs.

\paragraph{\textbf{LLM controller versus bandits and plug-in routers}}
Figure~\ref{fig:results_1} highlights the difference between an LLM-based router and the baselines visible through their model-selection patterns. K-NN, UniRoute, Smoothie, GreenServ, and RouteLLM never decode a routing token. They instead score embeddings,
clusters, or a small contextual vector, and then apply an explicit knob such as $\lambda$, $\alpha$, or a cost term.  For instance, GreenServ~$\lambda{=}0.5$ stays cheap while still mixing mid-size models, K-NN stays R1-heavy because neighbours vote for correctness ($418$ calls to R1-8B and $294$ to R1-32B), and RouteLLM mostly sends traffic to its strong arm.  In contrast, the LLM controller directly reads the question and generates a routing decision, enabling task-dependent model selection without a hand-built feature map. Under KL-GRPO, this results in structured routing, with MATH-500 favoring R1-8B, ARC favoring Gemma, and GSM8K favoring the $1.5$B model, which is the mix behind the even ARC and MATH-500 accuracies in Table~\ref{tab:accuracy_split}. KL-GRPO step~1500 distributes traffic across eight models and rarely selects R1-32B, while GRPO step~750 is more 7B-heavy and achieves lower energy at similar accuracy. Therefore, it can match UniRoute without large energy. GRPO step~750 is likewise mixed but more 7B-heavy, and therefore cheaper at similar accuracy. This flexibility can also lead to policy collapse. With GRPO~$\lambda_e{=}0.3$, the controller routes every item to Llama~3.1~8B, yielding zero selection entropy despite competitive aggregate energy and accuracy. SFT with $\lambda_e{=}0.5$ exhibits a different failure mode by over-selecting the $0.5$B model, resulting in low energy but poor accuracy. These results show that effective learned routing requires maintaining task-dependent model diversity rather than collapsing to a single low-cost decision.

\paragraph{\textbf{Latency and Power Profile}}
Latency and power follow the same three sections as accuracy. The high accuracy
rows draw about $121$--$175\,\mathrm{W}$ and take roughly $23$--$29\,\mathrm{s}$,
the middle rows draw about $48$--$85\,\mathrm{W}$ over $8$--$15\,\mathrm{s}$,
and the cheap rows draw about $23$--$47\,\mathrm{W}$ over $4$--$6\,\mathrm{s}$.
The TRL policies sit with Smoothie and GreenServ on power, at
$47$--$55\,\mathrm{W}$, rather than with K-NN, and KL-GRPO step~1500 is faster
than UniRoute while GRPO step~750 is faster than Smoothie. As before, the
duration column reflects answer model time only and does not include the
controller.

\paragraph{\textbf{Controller Energy and Training Cost}}

Tables report answer-model energy because that is the quantity every baseline
shares once a model is chosen.
Most competing methods are lookup-like mechanisms rather than generative controllers like the proposed router, so we
do not invent a controller cost for them or leave the comparison on a guess. To visualize detailed per-session power trajectories and investigate where the energy savings come from, 
Figure~\ref{fig:placeholder} and
Figure~\ref{fig:rl_vs_sft_power_timeseries_b01} (in Appendix) use the same four held-out
questions, the same $200\,\mathrm{ms}$ NVML protocol, and NVIDIA
A100-SXM4-$80$,GB GPUs. Across those sessions, the occupancy of the
board actively change. SFT remains the
longer decode (about $2.4$--$2.8\,\mathrm{s}$ versus about
$0.8$--$1.0\,\mathrm{s}$) because the warm-start samples emit more tokens, so
SFT controller energy stays higher even when both runs sit on the same GPU class. KL-GRPO controller energy is a few hundred joules in Fig.~\ref{fig:placeholder} and about $30\,\mathrm{J}$ in
Fig.~\ref{fig:rl_vs_sft_power_timeseries_b01}. Latency, not watts, is the stable signature of the training paradigm shift. That add-on is why we keep answer model energy in the main table.
That cost grows if the controller itself is scaled (larger routers, more candidates, or multi-turn routing) or if the pool moves to still heavier answer models ($48$B and above), where choosing \emph{which} call to issue
matters more than zeroing the router's own watts.
Depending on steps, total training cost varied from 0.32 to 0.40 kWh,
and remains a one-time investment for a reusable local controller.

\section{Conclusion and Future Works}

In this work, we look into the energy-efficient serving of LLMs, and investigate novel energy-aware model routing over a heterogeneous pool of LLMs via measured GPU energy rank. Instead of using model size or monetary cost as the routing objective, we propose to let the small language-model router evaluate the whole input queries' contexts and energy profiles and finetune the router model with tailored reward.  Such a router firstly undergoes supervised fine-tuning followed by reinforcement learning-based policy optimization, and it is evaluated on a held-out mixture of seven benchmarks using offline tournament logs. Our results show that routing policies can substantially alter the accuracy–energy trade-off by changing how frequently different answer models are selected, leading to a promising energy-efficient LLM serving paradigm while preserving model performance.
Compared with router trained via SFT, selected GRPO and KL-GRPO checkpoints achieve higher held-out accuracy while maintaining lower energy than several existing routing baselines at comparable operating points. 
These results suggest that just with a mild-size finetuning dataset collected, measured-energy-aware routing is effective for realizing heterogeneous LLM serving, which is both flexible and generalizable across task domains.
Future work will evaluate the approach in online and multi-turn agent settings and account for the full end-to-end routing overhead.
\label{sec:conclusion}

\bibliographystyle{acm}
\bibliography{bib}

\clearpage

\appendix
\section{Simulation Details}
\label{sec:appendix-sim-details}

Table~\ref{tab:train-test-split} lists how many questions we collected per
benchmark and how they split into train and held-out. We originally collected
$3{,}000$ tournament items. Before building the final evaluation set we dropped
$37$ questions where at least one pool model failed to load or return a usable
log, so those incomplete items never enter train or held-out scoring. The retained
mix is $2{,}050$ train and $913$ held-out questions. Train prefixes and held-out
suffixes are benchmark-specific as in the table. Every retained question was run
through all nine answer models to build the offline correctness and energy logs
used for oracle labeling, SFT, RL, and replay evaluation.

\begin{table}[h!]
\centering
\caption{Number of questions per benchmark used for training and held-out evaluation.
Collected counts are before removing $37$ incomplete tournament items.}
\label{tab:train-test-split}
\begin{tabular}{lccc}
\toprule
\textbf{Benchmark} & \textbf{Collected} & \textbf{Train} & \textbf{Held-out} \\
\midrule
MMLU           & 475 & 325 & 149 \\
GSM8K          & 473 & 325 & 144 \\
BBH            & 472 & 325 & 145 \\
ARC Challenge  & 472 & 325 & 147 \\
HellaSwag      & 472 & 325 & 147 \\
MATH-500       & 472 & 325 & 122 \\
HumanEval      & 164 & 100 &  59 \\
\midrule
\textbf{Total} & \textbf{3{,}000} & \textbf{2{,}050} & \textbf{913} \\
\bottomrule
\end{tabular}
\end{table}

\begin{figure*}[t]
\centering
\begin{minipage}[t]{0.48\textwidth}
\centering
\includegraphics[width=\linewidth,height=0.36\textheight,keepaspectratio]{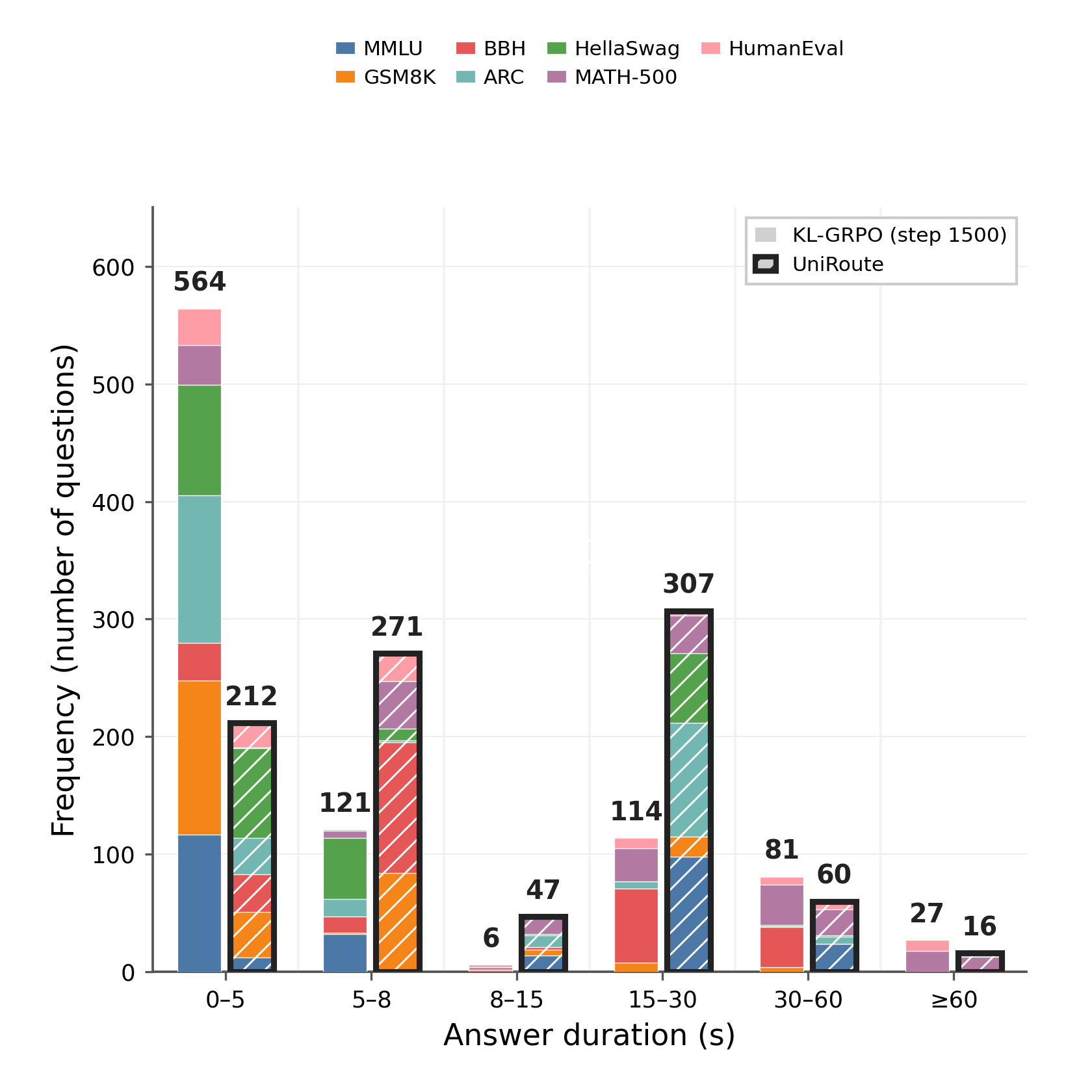}
\caption{Held-out answer-duration band counts by benchmark for KL-GRPO step~1500
and UniRoute.}
\label{fig:duration_band_frequency_by_benchmark_klgrpo_uniroute}
\end{minipage}
\hfill
\begin{minipage}[t]{0.48\textwidth}
\centering
\includegraphics[width=\linewidth,height=0.36\textheight,keepaspectratio]{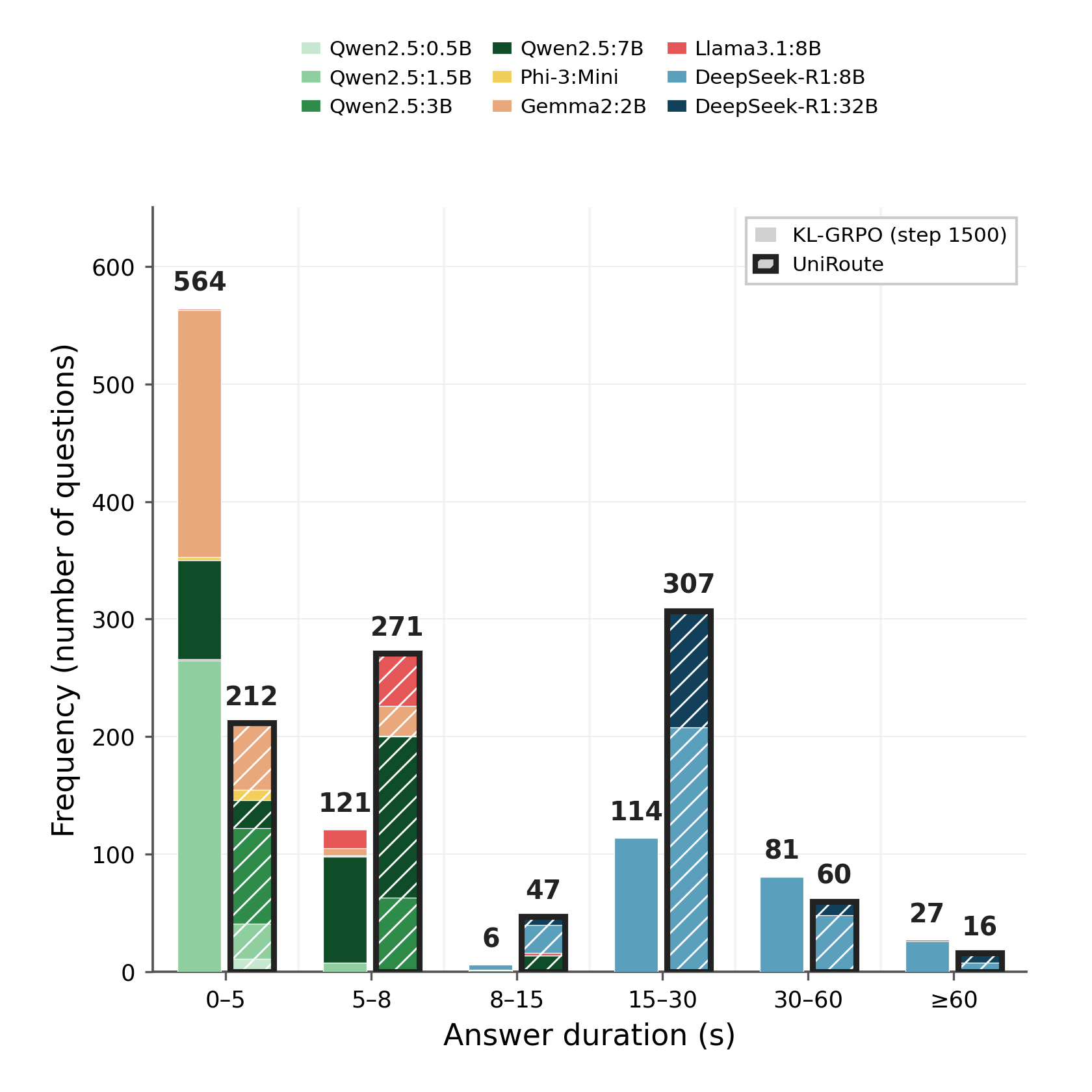}
\caption{Held-out answer-duration band counts aggregated over all benchmarks for
KL-GRPO step~1500 and UniRoute.}
\label{fig:duration_band_frequency_klgrpo_uniroute}
\end{minipage}
\end{figure*}

\section{Additional Simulation Results}
\label{sec:appendix-extra}

The duration plots explain why KL-GRPO step~1500 is cheaper than UniRoute at
almost the same held-out accuracy. Figure~\ref{fig:duration_band_frequency_klgrpo_uniroute}
shows that $564$ of $913$ KL-GRPO answers finish in $0$--$5\,\mathrm{s}$, versus
$212$ for UniRoute, while UniRoute places $307$ items in the $15$--$30\,\mathrm{s}$
band against $114$ for KL-GRPO. The stacks are model-specific. KL-GRPO fills the
short bins with Qwen2.5-1.5B, Gemma2-2B, and Qwen2.5-7B. UniRoute fills the
$15$--$30\,\mathrm{s}$ band with DeepSeek-R1-8B and DeepSeek-R1-32B. Neither policy
uses many calls above $60\,\mathrm{s}$ ($27$ versus $16$).
Figure~\ref{fig:duration_band_frequency_by_benchmark_klgrpo_uniroute} splits the
same bins by task. KL-GRPO's short bin is mostly GSM8K, ARC, and HellaSwag.
UniRoute's $5$--$8\,\mathrm{s}$ peak is BBH-heavy, and its $15$--$30\,\mathrm{s}$
peak mixes MMLU, ARC, and HellaSwag. MATH-500 and HumanEval dominate the
$30\,\mathrm{s}$ and longer bands for both policies, which is why a minority of
items still drive mean energy.

Figure~\ref{fig:router_trace_examples} shows two cases where the controller
picks a small correct arm instead of a large one. GSM8K q448 goes to
Qwen2.5-1.5B at $87.6\,\mathrm{J}$. MMLU q358 goes to Gemma2-2B at
$108.8\,\mathrm{J}$. The traces are illustrative. The written reasons are generic
and should not be read as causal explanations of the policy.

\begin{figure*}[t]
\centering
\footnotesize
\begin{minipage}[t]{0.48\textwidth}
\setlength{\fboxsep}{4pt}
\noindent\textbf{Example 1.} GSM8K, question~448.\\
A salesman bought $48$ sneakers for $\$576$. He sold $17$ at a Monday flash
sale for $\$20$ each and the remaining $31$ to a department store for $\$25$
each. What was his profit?
\vspace{0.3em}

\noindent\fcolorbox{green!50!black}{green!8}{%
\begin{minipage}{0.97\linewidth}
\textbf{Qwen2.5-1.5B} \textsc{(correct)} \hfill $87.6\,\mathrm{J}$\\[0.15em]
\textit{Controller.} \texttt{Choice: B}\\
\texttt{REASONING:} The chosen model (qwen2.5:1.5b) has been shown to perform well on tasks involving financial calculations.\\[0.15em]
\textit{Answer.} Cost per pair $\$576/48=\$12$. Flash sale $17\times\$20=\$340$.
Store $31\times\$25=\$775$. Profit $\$340+\$775-\$576=\$539$.\\[0.15em]
\textit{Final answer:} $539$
\end{minipage}}
\end{minipage}
\hfill
\begin{minipage}[t]{0.48\textwidth}
\setlength{\fboxsep}{4pt}
\noindent\textbf{Example 2.} MMLU, question~358.\\
Why does the sky appear blue? (multiple choice; correct option D: molecules
scatter blue light more effectively than red light.)
\vspace{0.3em}

\noindent\fcolorbox{green!50!black}{green!8}{%
\begin{minipage}{0.97\linewidth}
\textbf{Gemma2-2B} \textsc{(correct)} \hfill $108.8\,\mathrm{J}$\\[0.15em]
\textit{Controller.} \texttt{Choice: E}\\
\texttt{REASONING:} The chosen model (gemma2:2b) is known for its large language model capacity.\\[0.15em]
\textit{Answer.} Blue light is scattered more by air molecules than other colors.\\[0.15em]
\textit{Final answer:} D
\end{minipage}}
\end{minipage}
\caption{Two held-out routing examples. GSM8K q448 to Qwen2.5-1.5B ($87.6\,\mathrm{J}$).
MMLU q358 to Gemma2-2B ($108.8\,\mathrm{J}$). Controller text is from a free-form run.
Answers and joules are tournament logs.}
\label{fig:router_trace_examples}
\end{figure*}

\begin{figure*}[t]
\centering
\includegraphics[width=\textwidth,height=0.42\textheight,keepaspectratio]{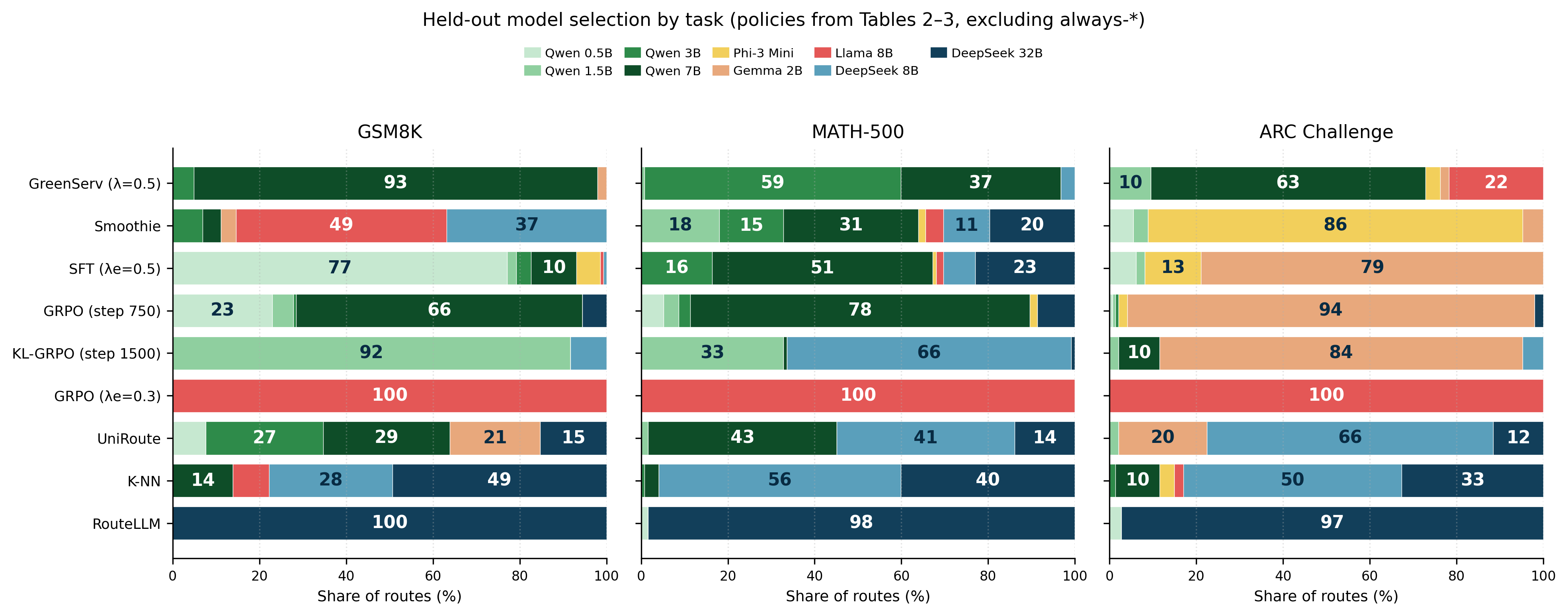}
\caption{Held-out model-selection mix on GSM8K, MATH-500, and ARC Challenge for
the non-always policies.
KL-GRPO step~1500 is mostly Qwen2.5-1.5B on GSM8K, DeepSeek-R1-8B on MATH-500,
and Gemma2-2B on ARC; GRPO step~750 is 7B-heavy on GSM8K/MATH and Gemma-heavy on
ARC; SFT still places mass on the $0.5$B model on GSM8K.
RouteLLM stays on R1-32B and GRPO~$\lambda_e{=}0.3$ locks onto Llama~3.1~8B on
every task.}
\label{fig:task_model_mix}
\end{figure*}

Figure~\ref{fig:task_model_mix} is the task-conditional view of the same mix.
KL-GRPO step~1500 uses Qwen2.5-1.5B on GSM8K, DeepSeek-R1-8B on MATH-500, and
Gemma2-2B on ARC. GRPO step~750 is 7B-heavy on GSM8K and MATH and Gemma-heavy on
ARC. SFT still puts mass on Qwen2.5-0.5B on GSM8K. RouteLLM stays on R1-32B.
GRPO $\lambda_e{=}0.3$ locks onto Llama~3.1~8B on every task. That is why
aggregate mid-band rows can hide collapse or under-routing.

Table~\ref{tab:trl_pareto_checkpoints} orders the TRL sweep by mean answer
energy. Matched $\lambda_e{=}0.3$ is cheap and collapsed ($57.3\%$ at
$297\,\mathrm{J}$). Matched $\lambda_e{=}0.7$ stays below $44\%$ accuracy.
The $0.5{\to}0.4$ family occupies the mid band. KL-GRPO step~1500 is the
highest-accuracy run in that family at $60.9\%$ and $1691.9\,\mathrm{J}$.
GRPO step~750 is the cheaper $0.5{\to}0.4$ operating point at $58.4\%$ and
$811.8\,\mathrm{J}$. Runs warmed from SFT ($\lambda_e{=}1.0$) jump to about
$74\%$ accuracy at more than $5\,\mathrm{kJ}$ because that warm start is the
accuracy-oracle policy, not a hybrid $\lambda_e{=}1.0$ label set.
Table~\ref{tab:trl_pareto_sft} makes the same split at SFT. Hybrid warm starts
sit near $45$--$54\%$ and $0.75$--$0.82\,\mathrm{kJ}$. SFT ($\lambda_e{=}1.0$)
is \texttt{sft\_acc} at $73.8\%$ and $4706.6\,\mathrm{J}$.

\begin{table*}[t]
\centering
\caption{TRL RL checkpoints from Table~\ref{tab:final_results} and the blue-circle sweep in Fig.~\ref{fig:pareto_accuracy_energy}, ordered by increasing mean energy.
Choice only means the controller is trained to emit a route letter
(\texttt{Choice: A}) without a free-form reasoning line.}
\label{tab:trl_pareto_checkpoints}
\small
\setlength{\tabcolsep}{3.0pt}
\renewcommand{\arraystretch}{0.95}
\begin{tabular}{lcccc}
\toprule
\textbf{Checkpoint} & \textbf{Acc.} & \textbf{Dur.\ (ms)} & \textbf{Energy (J)} & \textbf{Power (W)} \\
\midrule
GRPO step~1000 (SFT $\lambda_e{=}0.3$, RL $\lambda_e{=}0.3$) & 0.573 & 6002 & 297.0 & 46.8 \\
GRPO step~750 (SFT $\lambda_e{=}0.7$, RL $\lambda_e{=}0.7$, run2) & 0.421 & 5634 & 354.7 & 30.4 \\
GRPO step~1000 (SFT $\lambda_e{=}0.7$, RL $\lambda_e{=}0.7$, run3) & 0.432 & 6821 & 561.8 & 31.8 \\
GRPO step~1500 (SFT $\lambda_e{=}0.7$, RL $\lambda_e{=}0.7$) & 0.384 & 6070 & 569.3 & 27.9 \\
KL-GRPO step~500 (SFT $\lambda_e{=}0.7$, RL $\lambda_e{=}0.7$) & 0.527 & 7286 & 648.7 & 45.5 \\
KL-GRPO step~1536 (SFT $\lambda_e{=}0.3$, RL $\lambda_e{=}0.3$) & 0.503 & 7410 & 719.2 & 44.4 \\
KL-GRPO step~500 (SFT $\lambda_e{=}0.5$, RL $\lambda_e{=}0.4$, choice only) & 0.524 & 7987 & 786.6 & 46.0 \\
\textbf{GRPO step~750} (SFT $\lambda_e{=}0.5$, RL $\lambda_e{=}0.4$, choice only) & 0.584 & 7942 & 811.8 & 48.5 \\
KL-GRPO step~1750 (SFT $\lambda_e{=}0.5$, RL $\lambda_e{=}0.4$, choice only) & 0.563 & 8703 & 986.8 & 50.9 \\
KL-GRPO step~500 (SFT $\lambda_e{=}0.3$, RL $\lambda_e{=}0.3$) & 0.549 & 9814 & 1296.8 & 48.6 \\
\textbf{KL-GRPO step~500} (SFT $\lambda_e{=}0.5$, RL $\lambda_e{=}0.4$) & 0.591 & 10932 & 1339.8 & 47.1 \\
KL-GRPO step~500 (SFT $\lambda_e{=}0.2$, RL $\lambda_e{=}0.2$) & 0.552 & 11809 & 1526.8 & 49.8 \\
GRPO step~1500 (SFT $\lambda_e{=}0.5$, RL $\lambda_e{=}0.4$, choice only) & 0.581 & 11860 & 1588.5 & 54.8 \\
GRPO step~1000 (SFT $\lambda_e{=}0.5$, RL $\lambda_e{=}0.4$) & 0.575 & 12561 & 1599.5 & 49.2 \\
\textbf{KL-GRPO step~1500} (SFT $\lambda_e{=}0.5$, RL $\lambda_e{=}0.4$) & 0.609 & 13118 & 1691.9 & 55.2 \\
GRPO step~1750 (SFT $\lambda_e{=}0.5$, RL $\lambda_e{=}0.4$, choice only) & 0.589 & 11764 & 1796.2 & 78.9 \\
GRPO step~1250 (SFT $\lambda_e{=}0.2$, RL $\lambda_e{=}0.2$) & 0.590 & 14076 & 1875.9 & 52.9 \\
GRPO step~2000 (SFT $\lambda_e{=}1.0$, RL $\lambda_e{=}0.3$) & 0.744 & 25574 & 5033.7 & 170.9 \\
GRPO step~1250 (SFT $\lambda_e{=}1.0$, RL $\lambda_e{=}0.6$) & 0.736 & 25581 & 5051.1 & 173.5 \\
GRPO step~500 (SFT $\lambda_e{=}1.0$, RL $\lambda_e{=}1.0$) & 0.738 & 25825 & 5126.7 & 174.2 \\
\bottomrule
\end{tabular}
\end{table*}

\begin{table*}[t]
\centering
\caption{SFT warm-start checkpoints for the TRL sweep and Table~\ref{tab:final_results}.
The labelled SFT $\lambda_e{=}0.5$ marker in Fig.~\ref{fig:pareto_accuracy_energy} matches the $\lambda_e{=}0.5$ row.
SFT ($\lambda_e{=}1.0$) is the accuracy-oracle warm start \texttt{sft\_acc}, not a hybrid $\lambda_e{=}1.0$ label set.}
\label{tab:trl_pareto_sft}
\small
\setlength{\tabcolsep}{4pt}
\renewcommand{\arraystretch}{1.05}
\begin{tabular}{lcccc}
\toprule
\textbf{Checkpoint} & \textbf{Acc.} & \textbf{Dur.\ (ms)} & \textbf{Energy (J)} & \textbf{Power (W)} \\
\midrule
SFT ($\lambda_e{=}0.7$) & 0.468 & 7517 & 749.9 & 45.5 \\
SFT ($\lambda_e{=}0.3$) & 0.456 & 7588 & 776.7 & 45.0 \\
SFT ($\lambda_e{=}0.5$, choice only) & 0.544 & 7912 & 779.9 & 47.0 \\
SFT ($\lambda_e{=}0.5$) & 0.502 & 7821 & 822.8 & 45.9 \\
SFT ($\lambda_e{=}1.0$) & 0.738 & 24414 & 4706.6 & 160.3 \\
\bottomrule
\end{tabular}
\end{table*}

Table~\ref{tab:model_benchmark_profile} is the pool profile that those policies
draw from. On held-out testing, Always-Qwen2.5-7B is $64.0\%$ at $281.6\,\mathrm{J}$,
but BBH is only $26.2\%$, indicating Qwen2.5-7B's imbalanced performance across benchmarks. DeepSeek-R1-8B is the strongest overall arm
($79.1\%$), but also comes as the expensive always-medium baseline ($4127.9\,\mathrm{J}$).
DeepSeek-R1-32B is weaker than R1-8B on BBH ($27.6\%$ versus $69.7\%$) despite
higher energy. Train and held-out accuracy ranks are similar. Train energy is
higher because long MATH generations are more common in the train prefix.

\begin{table*}[t]
\centering
\footnotesize
\setlength{\tabcolsep}{2.4pt}
\renewcommand{\arraystretch}{0.88}
\caption{Held-out per-model accuracy (\%) and mean effective answer energy (J) on $913$ questions. Overall is item-weighted.}
\label{tab:model_benchmark_profile}
\resizebox{\textwidth}{!}{%
\begin{tabular}{llcccccccc}
\toprule
\textbf{Model} & & \textbf{MMLU} & \textbf{GSM8K} & \textbf{BBH} & \textbf{ARC} & \textbf{HellaSwag} & \textbf{MATH-500} & \textbf{HumanEval} & \textbf{Overall} \\
\midrule
Qwen 2.5 0.5B & Acc.
& 14.8 & 27.1 & 2.1 & 35.4 & 20.4 & 17.2 & 10.2 & 18.9 \\
& Energy
& 116.5 & 91.1 & 62.7 & 72.0 & 45.5 & 167.6 & 110.3 & 91.8 \\
\midrule
Qwen 2.5 1.5B & Acc.
& 37.6 & 59.0 & 8.3 & 68.7 & 44.9 & 32.8 & 30.5 & 41.4 \\
& Energy
& 126.5 & 129.7 & 84.0 & 92.9 & 66.1 & 226.7 & 105.0 & 117.1 \\
\midrule
Qwen 2.5 3B & Acc.
& 53.7 & 64.6 & 11.0 & 74.8 & 59.9 & 49.2 & 33.9 & 51.2 \\
& Energy
& 198.4 & 233.5 & 195.3 & 136.7 & 124.4 & 497.6 & 167.7 & 219.6 \\
\midrule
Phi-3 Mini 3.8B & Acc.
& 49.0 & 50.0 & 6.2 & 78.9 & 55.1 & 18.0 & 20.3 & 42.2 \\
& Energy
& 236.1 & 297.0 & 233.8 & 177.5 & 172.3 & 471.4 & 246.7 & 257.8 \\
\midrule
Gemma 2 2B & Acc.
& 38.3 & 55.6 & 10.3 & 74.8 & 56.5 & 20.5 & 25.4 & 42.2 \\
& Energy
& 108.5 & 141.3 & 100.6 & 63.2 & 46.4 & 227.1 & 92.1 & 109.9 \\
\midrule
Qwen 2.5 7B & Acc.
& 59.1 & 84.0 & 26.2 & 84.4 & 61.9 & 67.2 & 67.8 & 64.0 \\
& Energy
& 219.9 & 311.6 & 263.4 & 149.2 & 141.0 & 701.0 & 221.5 & 281.6 \\
\midrule
Llama 3.1 8B & Acc.
& 50.3 & 81.2 & 37.9 & 78.9 & 57.8 & 36.9 & 50.8 & 57.3 \\
& Energy
& 263.4 & 331.9 & 305.8 & 204.4 & 176.6 & 571.4 & 238.0 & 297.0 \\
\midrule
DeepSeek-R1 8B & Acc.
& 59.7 & 92.4 & 69.7 & 93.2 & 70.7 & 86.1 & 89.8 & 79.1 \\
& Energy
& 2997.6 & 3318.5 & 3626.6 & 2095.0 & 3744.1 & 8882.8 & 6379.1 & 4127.9 \\
\midrule
DeepSeek-R1 32B & Acc.
& 65.1 & 89.6 & 27.6 & 92.5 & 72.8 & 86.9 & 94.9 & 73.5 \\
& Energy
& 3874.6 & 2354.2 & 4403.3 & 3276.1 & 3750.8 & 9080.4 & 16475.5 & 5112.4 \\
\bottomrule
\end{tabular}}
\end{table*}

\begin{table*}[t]
\centering
\footnotesize
\setlength{\tabcolsep}{2.4pt}
\renewcommand{\arraystretch}{0.88}
\caption{Train per-model accuracy (\%) and mean effective answer energy (J) on $2{,}050$ questions. Overall is item-weighted.}
\label{tab:model_benchmark_profile_train}
\resizebox{\textwidth}{!}{%
\begin{tabular}{llcccccccc}
\toprule
\textbf{Model} & & \textbf{MMLU} & \textbf{GSM8K} & \textbf{BBH} & \textbf{ARC} & \textbf{HellaSwag} & \textbf{MATH-500} & \textbf{HumanEval} & \textbf{Overall} \\
\midrule
Qwen 2.5 0.5B & Acc.
& 20.9 & 34.5 & 6.8 & 29.5 & 29.5 & 11.7 & 30.0 & 22.5 \\
& Energy
& 125.8 & 300.7 & 859.4 & 63.4 & 57.6 & 2063.2 & 117.1 & 555.8 \\
\midrule
Qwen 2.5 1.5B & Acc.
& 42.5 & 61.5 & 18.8 & 65.5 & 40.0 & 32.0 & 64.0 & 44.4 \\
& Energy
& 151.3 & 888.4 & 803.3 & 80.4 & 72.9 & 2000.3 & 100.8 & 638.5 \\
\midrule
Qwen 2.5 3B & Acc.
& 45.8 & 52.0 & 23.7 & 71.7 & 56.6 & 46.2 & 52.0 & 49.5 \\
& Energy
& 203.6 & 250.4 & 698.4 & 129.4 & 135.1 & 1198.4 & 172.6 & 423.1 \\
\midrule
Phi-3 Mini 3.8B & Acc.
& 41.8 & 58.5 & 16.9 & 80.9 & 55.4 & 11.1 & 30.0 & 43.4 \\
& Energy
& 249.5 & 1806.7 & 4574.7 & 167.4 & 181.6 & 2971.3 & 238.2 & 1589.2 \\
\midrule
Gemma 2 2B & Acc.
& 40.9 & 51.7 & 21.2 & 71.1 & 53.5 & 14.8 & 45.0 & 42.3 \\
& Energy
& 112.5 & 153.5 & 92.3 & 53.1 & 56.2 & 1740.8 & 111.5 & 355.5 \\
\midrule
Qwen 2.5 7B & Acc.
& 59.4 & 86.2 & 24.6 & 88.3 & 64.6 & 59.4 & 70.0 & 64.0 \\
& Energy
& 273.4 & 338.7 & 241.0 & 137.0 & 159.4 & 1292.8 & 213.4 & 397.6 \\
\midrule
Llama 3.1 8B & Acc.
& 43.7 & 76.6 & 40.6 & 80.9 & 55.4 & 31.4 & 56.0 & 54.8 \\
& Energy
& 336.8 & 1242.2 & 1905.0 & 192.4 & 188.0 & 16106.0 & 223.4 & 3176.9 \\
\midrule
DeepSeek-R1 8B & Acc.
& 66.8 & 95.1 & 71.4 & 93.2 & 72.9 & 85.8 & 85.0 & 81.1 \\
& Energy
& 3812.9 & 3586.1 & 5815.8 & 2163.6 & 4468.3 & 13004.1 & 8476.1 & 5621.5 \\
\midrule
DeepSeek-R1 32B & Acc.
& 70.8 & 94.2 & 32.6 & 94.5 & 69.8 & 79.1 & 93.0 & 74.4 \\
& Energy
& 4981.8 & 2448.9 & 6337.2 & 3322.3 & 3862.0 & 12974.5 & 13168.3 & 6021.0 \\
\bottomrule
\end{tabular}}
\end{table*}

\begin{figure*}[t]
\centering
\includegraphics[width=\textwidth,height=0.48\textheight,keepaspectratio]{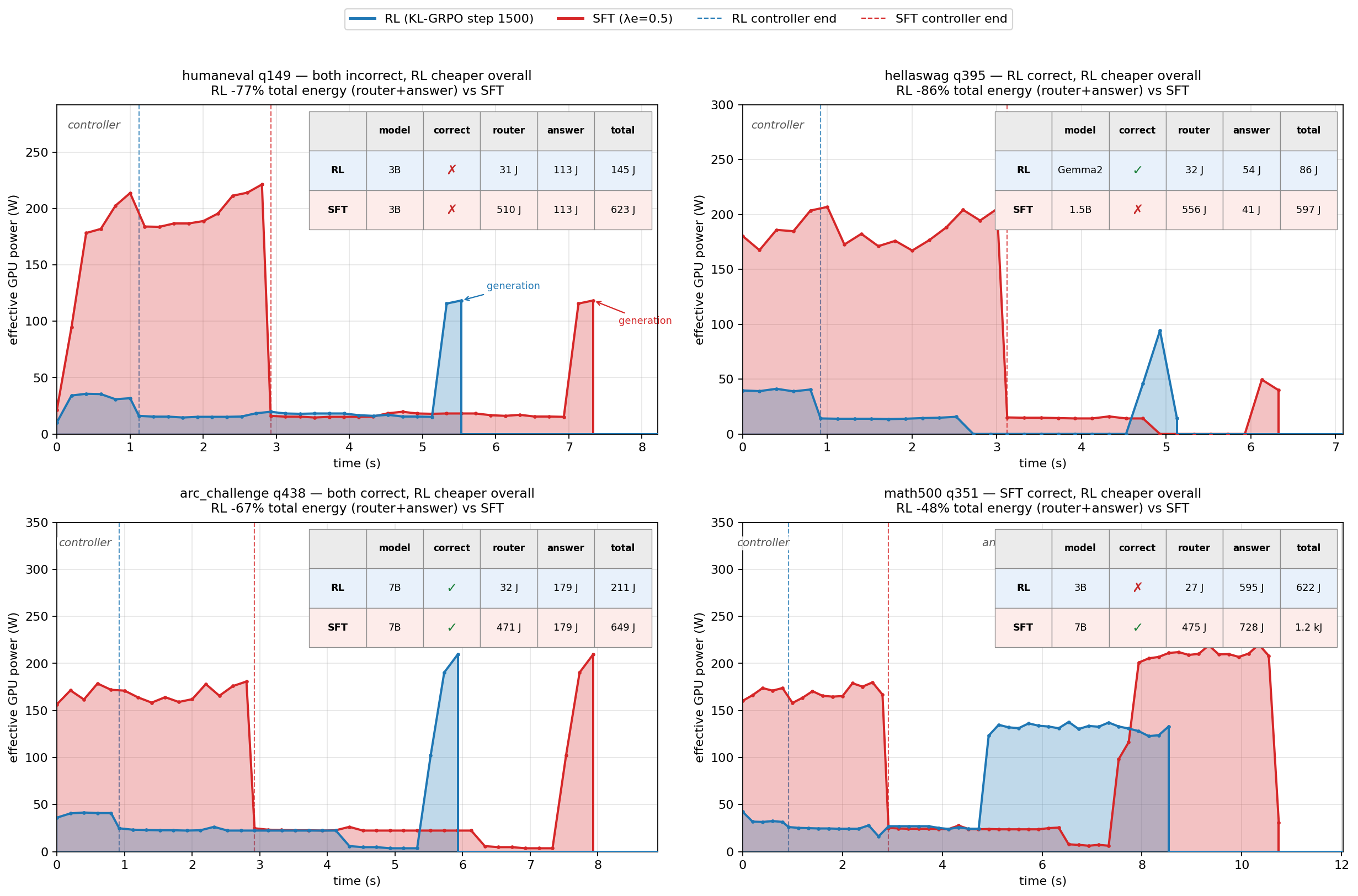}
\caption{Same four held-out questions as Fig.~\ref{fig:placeholder} on a later
NVIDIA A100-SXM4-$80$,GB session. KL-GRPO step~1500 of a different KL beta variation versus SFT. Router energy
is about $30\,\mathrm{J}$ for KL-GRPO and a few hundred joules for the longer
SFT decode. Answer curves reuse tournament logs.}
\label{fig:rl_vs_sft_power_timeseries_b01}
\end{figure*}

Figure~\ref{fig:rl_vs_sft_power_timeseries_b01} repeats the four-question
layout of Fig.~\ref{fig:placeholder} on a later A100-SXM4-$80$GB session using
the KL-GRPO step~1500 of a different KL beta variation and the matching
SFT eval. The KL-GRPO controller is still about $0.8\,\mathrm{s}$ and here
about $27$--$32\,\mathrm{J}$. SFT is still about $2.8\,\mathrm{s}$ and about
$470$--$560\,\mathrm{J}$. Answer traces for HumanEval q149 and ARC q438 overlap
because both policies pick the same model. The large SFT fill is extra
controller tokens, not a different GPU. The two figures disagree on absolute
watts because board occupancy differed across sessions on the same SKU.
Controller energy is therefore a mild 
add-on to Table~\ref{tab:final_results}, not a reason to move the main
comparison off answer joules.

\end{document}